\documentstyle[manuscript,aps,prl]{revtex}
\begin{document}
%\draft

\onecolumn

\title{Multi-parameter Entanglement in Quantum Interferometry}
\author{Mete~Atat\"{u}re,$^1$ Giovanni~Di~Giuseppe,$^2$ Matthew~D.~ Shaw,$^2$ Alexander~V.~Sergienko,$^{1,2}$
Bahaa~E.~A.~Saleh,$^2$ and Malvin~C.~Teich$^{1,2}$}
\address{Quantum Imaging Laboratory,\\ $^1$Department of Physics and
$^2$Department of Electrical and Computer Engineering,\\ Boston
University,
8 Saint Mary's Street, Boston, MA 02215}

\date{\today}
\maketitle
\begin{abstract}

The role of multi-parameter entanglement in quantum interference
from collinear type-II spontaneous parametric down-conversion is
explored using a variety of aperture shapes and sizes, in regimes
of both ultrafast and continuous-wave pumping. We have developed
and experimentally verified a theory of down-conversion which
considers a quantum state that can be concurrently entangled in
frequency, wavevector, and polarization. In particular, we
demonstrate deviations from the familiar triangular interference
dip, such as asymmetry and peaking. These findings improve our
capacity to control the quantum state produced by spontaneous
parametric down-conversion, and should prove useful to those
pursuing the many proposed applications of down-converted light.

\end{abstract}


\tableofcontents \pagebreak

\section{Introduction}

In the nonlinear-optical process of spontaneous parametric
down-conversion (SPDC) \cite{SPDC}, in which a laser beam
illuminates a nonlinear-optical crystal, pairs of photons are
generated in a state that can be entangled \cite{Schrodinger}
concurrently in frequency, momentum, and polarization. A
significant number of experimental efforts designed to verify the
entangled nature of such states have been carried out on states
entangled in a {\em single} parameter, such as in energy
\cite{Energy}, momentum \cite{Momentum}, or polarization
\cite{Polarization}. In general, the quantum state produced by
SPDC is not factorizable into independently entangled
single-parameter functions. Consequently any attempt to access one
parameter is affected by the presence of the others. A common
approach to quantum interferometry to date has been to choose a
single entangled parameter of interest and eliminate the
dependence of the quantum state on all other parameters. For
example, when investigating polarization entanglement, spectral
and spatial filtering are typically imposed in an attempt to
restrict attention to polarization  alone.

A more general approach to this problem is to consider and exploit
the concurrent entanglement from the outset. In this approach, the
observed quantum-interference pattern in one parameter, such as
polarization, can be modified at will by controlling the
dependence of the state on the other parameters, such as frequency
and transverse wavevector. This strong interdependence has its
origin in the nonfactorizability of the quantum state into product
functions of the separate parameters.

In this paper we theoretically and experimentally study how the
polarization quantum-interference pattern, presented as a function
of relative temporal delay between the photons of an entangled
pair, is modified by controlling the optical system through
different kinds of spatial apertures. The effect of the spectral
profile of the pump field is investigated by using both a
continuous-wave and a pulsed laser to generate SPDC. The role of
the spatial profile of the pump field is also studied
experimentally by restricting the pump-beam diameter at the face
of the nonlinear crystal.

Spatial effects in Type-I SPDC have previously been investigated,
typically in the context of imaging with spatially resolving
detection systems~\cite{Angle-Spread}. The theoretical formalism
presented here for Type-II SPDC is suitable for extension to
Type-I in the presence of an arbitrary optical system and
detection apparatus. Our study leads to a deeper physical
understanding of multi-parameter entangled two-photon states and
concomitantly provides a route for engineering these states for
specific applications, including quantum information processing.

\section{Multi-parameter Entangled-State Formalism}

In this section we present a multidimensional analysis of the
entangled-photon state generated via type-II SPDC. To admit a
broad range of possible experimental schemes we consider, in turn,
the three distinct stages of any experimental apparatus: the
generation, propagation, and detection of the quantum state
\cite{Duality}.

\subsection{Generation}

By virtue of the relatively weak interaction in the nonlinear
crystal, we consider the two-photon state generated within the
confines of first-order time-dependent perturbation theory:

\begin{equation}
\label{Psi-Definition} | \Psi^{(2)}\rangle ~ \sim ~{{\rm
i}\over{\hbar}}\int\limits_{\,\,\, t_{0}}^{ \,\,\, t}dt'~\hat
H_{\rm int}(t')~|0\rangle \,.
\end{equation}

\noindent Here $\hat H_{\rm int}(t')$ is the interaction
Hamiltonian, $[t_{0},t]$ is the duration of the interaction, and
$|0\rangle$ is the initial vacuum state.  The interaction
Hamiltonian governing this phenomenon is \cite{Biphoton}

\begin{equation}
\label{Hamiltonian} \hat H_{\rm int}(t') \sim
\chi^{(2)}\int\limits_{\,\,\,\,\,\, {V}}d{\bf r}~\hat
E_{p}^{(+)}({\bf r},t')\,\hat E_{o}^{(-)} ({\bf r},t')\,\hat
E_{e}^{(-)}({\bf r},t')~+~{\rm H.c.}\,,
\end{equation}

\noindent where $\chi^{(2)}$ is the second-order susceptibility
and $V$ is the volume of the nonlinear medium in which the
interaction takes place. The operator $\hat E_{j}^{(\pm)}({\bf
r},t')$ represents the positive- (negative-) frequency portion of
the $j$th electric-field operator, with the subscript $j$
representing the pump ($p$), ordinary ($o$), and extraordinary
($e$) waves at position ${\bf r}$ and time $t'$, and {\rm H.c.}
stands for Hermitian conjugate.  Because of the high intensity of
the pump field it can be represented by a classical c-number,
rather than as an operator, with an arbitrary spatiotemporal
profile given by

\begin{equation}
\label{Pump-3DGeneral}
E_{p}({\bf r},t)=\int d{\bf k}_{p}~{\tilde E}_{p}({\bf k}_{p})e^{{\rm
i}{\bf k}_{p} \cdot {\bf r}}e^{{\rm -i}\omega_{p}({\bf k}_{p})t}\,,
\end{equation}

\noindent where ${\tilde E}_{p}({\bf k}_{p})$ is the
complex-amplitude profile of the field as a function of the
wavevector ${\bf k}_{p}$.

We decompose the three-dimensional wavevector ${\bf k}_{p}$ into a
two-dimensional transverse wavevector ${\bf q}_{p}$ and frequency
$\omega_{p}$, so that Eq.~(\ref{Pump-3DGeneral}) takes the form

\begin{equation}
\label{Pump-General}
E_{p}({\bf r},t)=\int d{\bf q}_{p}\,d\omega_{p}~{\tilde
  E}_{p}({\bf q}_{p};\omega_{p})e^{{\rm i}\kappa_{p}z}e^{{\rm i}{\bf q}_{p} \cdot {\bf x}}e^{{\rm -i}\omega_{p}t}\,,
\end{equation}

\noindent where ${\bf x}$ spans the transverse plane perpendicular
to the propagation direction $z$.  In a similar way the ordinary
and extraordinary fields can be expressed in terms of the
quantum-mechanical creation operators $\hat a^{\dagger}({\bf
q},\omega)$ for the $({\bf q},\omega)$ modes as

\begin{equation}
\label{Field-Generation} \hat E_{j}^{(-)}({\bf r},t)=\int d{\bf
q}_{j}\,d\omega_{j}~e^{{\rm -i}\kappa_{j}z}e^{{\rm -i}{\bf q}_{j}
\cdot {\bf x}}e^{{\rm i}\omega_{j}t}\,\hat a^{\dagger}_{j}({\bf
q}_{j},\omega_{j})\,,
\end{equation}

\noindent where the subscript $j=o,e$.  The longitudinal component
of ${\bf k}$, denoted $\kappa$, can be written in terms of the
$({\bf q},\omega)$ pair as \cite{Duality,Decompose}

\begin{equation}
\label{Kappa-General} \kappa=\sqrt{\left[n(\omega,\theta)\,\omega
\over c\right]^{2}-|{\bf q}|^2}\,,
\end{equation}

\noindent where $c$ is the speed of light in vacuum, $\theta$ is
the angle between  ${\bf k}$ and the optical axis of the nonlinear
crystal (see Fig.~1), and $n(\omega,\theta)$ is the index of
refraction in the nonlinear medium. Note that the symbol
$n(\omega,\theta)$ in Eq. (\ref{Kappa-General}) represents the
extraordinary refractive index $n_{e}(\omega,\theta)$ when
calculating $\kappa$ for extraordinary waves, and the ordinary
refractive index $n_{o}(\omega)$ for ordinary waves.

Substituting Eqs. (\ref{Pump-General}) and
(\ref{Field-Generation}) into Eqs. (\ref{Psi-Definition}) and
(\ref{Hamiltonian}) yields the quantum state at the output of the
nonlinear crystal:

\begin{equation}
\label{Psi-General} |\Psi^{(2)}\rangle \sim \int d{\bf q}_{o}d{\bf
q}_{e}\,d\omega_{o}d\omega_{e} ~\Phi({\bf q}_{o},{\bf
q}_{e};\omega_{o},\omega_{e})\hat a^{\dagger}_{o}({\bf
q}_{o},\omega_{o})\hat a^{\dagger}_{e}({\bf
q}_{e},\omega_{e})|0\rangle\,,
\end{equation}

\noindent with

\begin{equation}
\label{Phi-General} \Phi({\bf q}_{o},{\bf
q}_{e};\omega_{o},\omega_{e})~=~{\tilde E}_{p}({\bf q}_{o}+{\bf
q}_{e};\omega_{o}+\omega_{e})\, L\,{\rm sinc}\left({{L\Delta}\over
2}\right)e^{{\rm -i}{{L\Delta}\over2}}\,,
\end{equation}

\noindent where $L$ is the thickness of the crystal and $\Delta =
\kappa_{p}-\kappa_{o}-\kappa_{e}$ where $\kappa_{j}$ ($j=p, o, e$)
is related to the indices $({\bf q}_j,\omega_j)$ via relations
similar to Eq.~(\ref{Kappa-General}).  The nonseparability of the
function $\Phi({\bf q}_{o},{\bf q}_{e};\omega_{o},\omega_{e})$ in
Eqs. (\ref{Psi-General}) and (\ref{Phi-General}), recalling
(\ref{Kappa-General}), is the hallmark of {\em concurrent}
multi-parameter entanglement.

\subsection{Propagation}

Propagation of the down-converted light between the planes of
generation and detection is characterized by the transfer function
of the optical system. The biphoton probability amplitude
\cite{Biphoton} at the space-time coordinates $({\bf
x}_{A},t_{A})$ and $({\bf x}_{B},t_{B})$ where detection takes
place is defined by
\begin{equation}
\label{Biphoton-Definition} A({\bf x}_{A},{\bf
x}_{B};t_{A},t_{B})=\langle0| \hat E_{A}^{(+)}({\bf
x}_{A},t_{A})\hat E_{B}^{(+)}({\bf x}_{B},t_{B})
|\Psi^{(2)}\rangle \,.
\end{equation}

\noindent The explicit forms of the quantum operators at the
detection locations are represented by \cite{Duality}

\begin{eqnarray}
\label{Field-Detector-General} \hat E_{A}^{(+)}({\bf
x}_{A},t_{A})=\int  d{\bf q}\,d\omega~e^{{\rm -i}\omega
t_{A}}\left[{\cal H}_{Ae}({\bf x}_{A},{\bf q};\omega)\hat
a_{e}({\bf q},\omega)+{\cal H}_{Ao}({\bf x}_{A},{\bf
q};\omega)\hat a_{o}({\bf q},\omega)\right]\,,\nonumber\\\hat
E_{B}^{(+)}({\bf x}_{B},t_{B})=\int d{\bf q}\,d\omega~e^{{\rm
-i}\omega t_{B}}\left[{\cal H}_{Be}({\bf x}_{B},{\bf
q};\omega)\hat a_{e}({\bf q},\omega)+{\cal H}_{Bo}({\bf
x}_{B},{\bf q};\omega)\hat a_{o}({\bf q},\omega)\right]\,,
\end{eqnarray}

\noindent where the transfer function ${\cal H}_{ij}$ ($i=A,B$ and
$j=e,o$) describes the propagation of a $({\bf q},\omega)$ mode
from the nonlinear-crystal output plane to the detection plane.
Substituting Eqs. (\ref{Psi-General}) and
(\ref{Field-Detector-General}) into
Eq.~(\ref{Biphoton-Definition}) yields a general form for the
biphoton probability amplitude:

\begin{eqnarray}
\label{Biphoton-General} A({\bf x}_{A},{\bf
x}_{B};t_{A},t_{B})&=\int d{\bf q}_{o}d{\bf
q}_{e}\,d\omega_{o}d\omega_{e}&~\Phi({\bf q}_{o},{\bf
q}_{e};\omega_{o},\omega_{e})\nonumber\\&&\times \left[{\cal
H}_{Ae}({\bf x}_{A},{\bf q}_{e};\omega_{e}){\cal H}_{Bo}({\bf
x}_{B},{\bf q}_{o};\omega_{o})\,e^{{\rm
-i}(\omega_{e}t_{A}+\omega_{o}t_{B})}\right.\nonumber\\&&\left.~~~+
{\cal H}_{Ao}({\bf x}_{A},{\bf q}_{o};\omega_{o}){\cal
H}_{Be}({\bf x}_{B},{\bf q}_{e};\omega_{e})\,e^{{\rm
-i}(\omega_{o}t_{A}+\omega_{e}t_{B})}\right].
\end{eqnarray}

\noindent By choosing optical systems with explicit forms of the
functions ${\cal H}_{Ae}$, ${\cal H}_{Ao}$, ${\cal H}_{Be}$, and
${\cal H}_{Bo}$, the overall biphoton probability amplitude can be
constructed as desired.

\subsection{Detection}

The formulation of the detection process requires some knowledge
of the detection apparatus. Slow detectors, for example, impart
temporal integration while detectors of finite area impart spatial
integration. One extreme case is realized when the temporal
response of a {\em point} detector is spread negligibly with
respect to the characteristic time scale of SPDC, namely the
inverse of down-conversion bandwidth. In this limit the
coincidence rate reduces to

\begin{equation}
\label{Coincidence-Fast}
R = |A({\bf x}_{A},{\bf x}_{B};t_{A},t_{B})|^2\,.
\end{equation}

\noindent On the other hand, quantum-interference experiments
typically make use of slow {\em bucket} detectors. Under these
conditions, the coincidence count rate $R$ is readily expressed in
terms of the biphoton probability amplitude as

\begin{equation}
\label{Coincidence-Slow}
R = \int d{\bf x}_{A}d{\bf x}_{B}\,dt_{A}dt_{B}~|A({\bf x}_{A},{\bf x}_{B};t_{A},t_{B})|^2\,.
\end{equation}

\section{Multi-parameter Entangled-State Manipulation}

In this section we apply the mathematical description presented
above to specific configurations of a quantum interferometer.
Since the evolution of the state is ultimately described by the
transfer function ${\cal H}_{ij}$, an explicit form of this
function is needed for each configuration of interest. Almost all
quantum-interference experiments performed to date have a common
feature, namely that the transfer function ${\cal H}_{ij}$ in
Eq.~(\ref{Biphoton-General}), with $i=A,B$ and $j=o,e$, can be
separated into diffraction-dependent and -independent terms as

\begin{equation}
\label{H-Separable} {\cal H}_{ij}({\bf x}_{i},{\bf
q};\omega)~=~{\cal T}_{ij}\,H({\bf x}_{i},{\bf q};\omega)\,,
\end{equation}

\noindent where the diffraction-dependent terms are grouped in $H$
and the diffraction-independent terms are grouped in ${\cal
T}_{ij}$ (see Fig.~2). Free space, apertures, and lenses, for
example, can be treated as diffraction-dependent elements while
beam splitters, temporal delays, and waveplates can be considered
as diffraction-independent elements. For collinear SPDC
configurations, for example, in the presence of a relative
optical-path delay $\tau$ between the ordinary and the
extraordinary polarized photons, as illustrated in Fig.~3(a),
${\cal T}_{ij}$ is simply

\begin{equation}
\label{Tab-1}
{\cal T}_{ij}=({\bf e}_{i}\cdot {\bf e}_{j})\,e^{{\rm -i}\omega\tau\delta_{ej}}\,,
\end{equation}

\noindent where the symbol $\delta_{ej}$ is the Kronecker delta
with $\delta_{ee}=1$ and $\delta_{eo}=0$. The unit vector ${\bf
e}_{i}$ describes the orientation of each polarization analyzer in
the experimental apparatus, while ${\bf e}_{j}$ is the unit vector
that describes the polarization of each down-converted photon.

Using the expression for ${\cal H}_{ij}$ given in
Eq.~(\ref{H-Separable}) in the general biphoton probability
amplitude given in Eq.~(\ref{Biphoton-General}), we construct a
compact expression for all systems that can be separated into
diffraction-dependent and -independent elements as:

\begin{eqnarray}
\label{Biphoton-Separable-Definition} A({\bf x}_{A},{\bf
x}_{B};t_{A},t_{B})&=\int d{\bf q}_{o}d{\bf
q}_{e}\,d\omega_{o}d\omega_{e}&~\Phi({\bf q}_{o},{\bf
q}_{e};\omega_{o},\omega_{e})\nonumber\\&&\times \left[{\cal
T}_{Ae}H({\bf x}_{A},{\bf q}_{e};\omega_{e}){\cal T}_{Bo}H({\bf
x}_{B},{\bf q}_{o};\omega_{o})\,e^{{\rm
-i}(\omega_{e}t_{A}+\omega_{o}t_{B})}\right.\nonumber\\&&\left.~~~+{\cal
T}_{Ao}H({\bf x}_{A},{\bf q}_{o};\omega_{o}){\cal T}_{Be}H({\bf
x}_{B},{\bf q}_{e};\omega_{e})\,e^{{\rm
-i}(\omega_{o}t_{A}+\omega_{e}t_{B})}\right]\,.
\end{eqnarray}

\noindent Given the general form of the biphoton probability
amplitude for a separable system, we now proceed to investigate
several specific experimental arrangements. For the experimental
work presented in this paper the angle between ${\bf e}_{i}$ and
${\bf e}_{j}$ is $45^{\circ}$, so ${\cal T}_{ij}$ can be
simplified by using $({\bf e}_{i}\cdot {\bf e}_{j})=\pm {1 \over
\sqrt{2}}$ \cite{Polarization}.  Substituting this into
Eq.~(\ref{H-Separable}), the biphoton probability amplitude
becomes

\begin{eqnarray}
\label{Biphoton-Separable} A({\bf x}_{A},{\bf
x}_{B};t_{A},t_{B})&=\int d{\bf q}_{o}d{\bf
q}_{e}\,d\omega_{o}d\omega_{e}&~{\tilde E}_{p}({\bf q}_{o}+{\bf
q}_{e};\omega_{o}+\omega_{e})\, L\,{\rm sinc}\left({{L\Delta}\over
2}\right)e^{{\rm -i}{{L\Delta}\over2}}\,e^{{\rm
-i}\omega_{e}\tau}\nonumber\\&&\times \left[\,H({\bf x}_{A},{\bf
q}_{e};\omega_{e})H({\bf x}_{B},{\bf q}_{o};\omega_{o})\,e^{{\rm
-i}(\omega_{e}t_{A}+\omega_{o}t_{B})}\right.\nonumber\\&&\left.~~~-H({\bf
x}_{A},{\bf q}_{o};\omega_{o})H({\bf x}_{B},{\bf
q}_{e};\omega_{e})\,e^{{\rm
-i}(\omega_{o}t_{A}+\omega_{e}t_{B})}\right]\,.
\end{eqnarray}

\noindent Substitution of Eq.~(\ref{Biphoton-Separable}) into
Eq.~(\ref{Coincidence-Slow}) gives an expression for the
coincidence-count rate given an arbitrary pump profile and optical
system.

However, this expression is unwieldy for purposes of predicting
interference patterns except in certain cases, where integration
can be swiftly performed. In particular, we consider three cases
of the spatial and spectral profile of the pump field in
Eq.~(\ref{Biphoton-Separable}): that of a polychromatic planewave,
that of a monochromatic planewave, and that of a monochromatic
beam with arbitrary spatial profile. These various cases will be
used subsequently for cw and pulse-pumped SPDC studies.

First, a non-monochromatic planewave pump field is described
mathematically by

\begin{equation}
\label{Pump-pulse} {\tilde E}_{p}({\bf
q}_{p};\omega_{p})\,=\,\delta({\bf q}_{p}){\cal
E}_p(\omega_{p}-\omega_{p}^{0})\,,
\end{equation}

\noindent where ${\cal E}_p(\omega_{p}-\omega_{p}^{0})$ is the
spectral profile of the pump field.
Equation~(\ref{Biphoton-Separable}) then takes the form

\begin{eqnarray}
\label{Biphoton-Separable-Pulse} A({\bf x}_{A},{\bf
x}_{B};t_{A},t_{B})&=\int d{\bf q}\,d\omega_{o}d\omega_{e}\,&
{\cal E}_p(\omega_{o}+\omega_{e}-\omega_{p}^{0})\,L\,{\rm
sinc}\left({{L\Delta}\over 2}\right)e^{{\rm
-i}{{L\Delta}\over2}}\,e^{{\rm
-i}\omega_{e}\tau}\,\nonumber\\&&\times \left[\,H({\bf x}_{A},{\bf
q};\omega_{e})H({\bf x}_{B},-{\bf q};\omega_{o})\,e^{{\rm
-i}(\omega_{e}t_{A}+\omega_{o}t_{B})}\right.\nonumber\\&&\left.~~~-H({\bf
x}_{A},-{\bf q};\omega_{o})H({\bf x}_{B},{\bf
q};\omega_{e})\,e^{{\rm
-i}(\omega_{o}t_{A}+\omega_{e}t_{B})}\right]\,.
\end{eqnarray}

Second, a monochromatic planewave pump field is described by

\begin{equation}
\label{Pump-CW} {\tilde E}_{p}({\bf
q}_{p};\omega_{p})\,=\,\delta({\bf
q}_{p})\delta(\omega_{p}-\omega_{p}^{0})\,,
\end{equation}

\noindent whereupon Eq.~(\ref{Biphoton-Separable-Pulse}) becomes

\begin{eqnarray}
\label{Biphoton-Separable-1} A({\bf x}_{A},{\bf
x}_{B};t_{A},t_{B})&=\int d{\bf q}\,d\omega\,& L\,{\rm
sinc}\left({{L\Delta}\over 2}\right)e^{{\rm
-i}{{L\Delta}\over2}}\,e^{{\rm -i}\omega\tau}\,e^{{\rm
-i}\omega_{p}^{0}(t_{A}+t_{B})}\nonumber\\&&\times \left[\,H({\bf
x}_{A},{\bf q};\omega)H({\bf x}_{B},-{\bf
q};\omega_{p}^{0}-\omega)\,e^{{\rm
-i}\omega(t_{A}-t_{B})}\right.\nonumber\\&&\left.~~~-H({\bf
x}_{A},-{\bf q};\omega_{p}^{0}-\omega)H({\bf x}_{B},{\bf
q};\omega)\,e^{{\rm i}\omega(t_{A}-t_{B})}\right]\,.
\end{eqnarray}

\noindent In this case the nonfactorizability of the state is due
solely to the phase matching.

Third, we examine the effect of the spatial distribution of the
pump field by considering a monochromatic field with an arbitrary
spatial profile described by

\begin{equation}
\label{Pump-CW} {\tilde E}_{p}({\bf
q}_{p};\omega_{p})\,=\,\Gamma({\bf
q}_{p})\delta(\omega_{p}-\omega_{p}^{0})\,,
\end{equation}

\noindent where $\Gamma({\bf q}_{p})$ characterizes the spatial
profile of the pump field through transverse wavevectors. In this
case Eq.~(\ref{Biphoton-Separable}) simplifies to

\begin{eqnarray}
\label{Biphoton-Separable-Spatial} A({\bf x}_{A},{\bf
x}_{B};t_{A},t_{B})&=\int d{\bf q}_{o}d{\bf q}_{e}\,d\omega\,&
\Gamma({\bf q}_{e}+{\bf q}_{o})\,L\,{\rm
sinc}\left({{L\Delta}\over 2}\right)e^{{\rm
-i}{{L\Delta}\over2}}\,e^{{\rm -i}\omega\tau}\,e^{{\rm
-i}\omega_{p}^{0}(t_{A}+t_{B})}\nonumber\\&&\times \left[\,H({\bf
x}_{A},{\bf q}_{e};\omega)H({\bf x}_{B},{\bf
q}_{o};\omega_{p}^{0}-\omega)\,e^{{\rm
-i}\omega(t_{A}-t_{B})}\right.\nonumber\\&&\left.~~~-H({\bf
x}_{A},{\bf q}_{o};\omega_{p}^{0}-\omega)H({\bf x}_{B},{\bf
q}_{e};\omega)\,e^{{\rm i}\omega(t_{A}-t_{B})}\right]\,.
\end{eqnarray}

Using Eq.~(\ref{Biphoton-Separable-1}) as the biphoton probability
amplitude for cw-pumped SPDC and Eq.
(\ref{Biphoton-Separable-Pulse}) for pulse-pumped SPDC in
Eq.~(\ref{Coincidence-Slow}), we can now investigate the behavior
of the quantum-interference pattern for optical systems described
by specific transfer functions $H$.
Eq.~(\ref{Biphoton-Separable-Spatial}) will be considered in
Section III.(C) to investigate the limit where the pump spatial
profile has a considerable effect on the quantum-interference
pattern.

The diffraction-dependent elements in most of these experimental
arrangements are illustrated in Fig.~3(b). To describe this system
mathematically via the function $H$, we need to derive the impulse
response function, also known as the point-spread function for
optical systems. A typical aperture diameter of $b=1$ cm at a
distance $d=1$ m from the crystal output plane yields
$b^{4}/4\lambda d^{3} < 10^{-2}$ using $\lambda=0.5$ $\mu$m,
guaranteeing the validity of the Fresnel approximation. We
therefore proceed with the calculation of the impulse response
function in this approximation. Without loss of generality we now
present a two-dimensional (one longitudinal and one transverse)
analysis of the impulse response function, extension to three
dimensions being straightforward.

Referring to Fig.~3(b), the overall propagation through this
system is broken into free-space propagation from the
nonlinear-crystal output surface $(x,0)$ to the plane of the
aperture $(x',d_{1})$, free-space propagation from the aperture
plane to the thin lens $(x'',d_{1}+d_{2})$, and finally free-space
propagation from the lens to the plane of detection
$(x_{i},d_{1}+d_{2}+f)$, with $i=A,B$. Free-space propagation of a
monochromatic spherical wave with frequency $\omega$ from $(x,0)$
to $(x',d_{1})$ over a distance $r$ is

\begin{equation}
\label{Free-Space-Propagation} e^{{\rm i}{\omega \over
c}r}=e^{{\rm i}{{\omega \over c}\sqrt{d_{1}^2+(x-x')^2}}} \approx
e^{{\rm i}{\omega \over c}d_{1}}e^{{\rm i}{\omega \over
2cd_{1}}(x-x')^2}\,.
\end{equation}

\noindent The spectral filter is represented mathematically by a
function ${\cal F}(\omega)$ and the aperture is represented by the
function $p(x)$. In the $(x',d_{1})$ plane, at the location of the
aperture, the impulse response function of the optical system
between planes $x$ and $x^\prime$ takes the form

\begin{equation}
\label{Impulse-Response-1} h(x',x;\omega) =  {\cal
F}(\omega)\,p(x')e^{{\rm i}{\omega \over c}d_{1}}e^{{\rm i}{\omega
\over 2cd_{1}}(x-x')^2}.
\end{equation}

\noindent Also, the impulse response function for the single-lens
system from the plane $(x',d_{1})$ to the plane
$(x_{i},d_{1}+d_{2}+f)$, as shown in Fig.~3(b), is

\begin{equation}
\label{Impulse-Response-Lens} h(x_{i},x';\omega)=e^{{\rm i}{\omega
\over c}(d_{2}+f)}e^{{\rm -i}{\omega x_{i}^2 \over
2cf}\left[{d_{2}\over f}-1\right]}e^{{\rm -i}{\omega x_{i}x'\over
cf}}\,.
\end{equation}

\noindent Combining this with Eq.~(\ref{Impulse-Response-1})
provides

\begin{equation}
\label{Impulse-Response-Total} h(x_{i},x;\omega)={\cal
F}(\omega)\,e^{{\rm i}{\omega \over c}(d_{1}+d_{2}+f)}e^{{\rm
-i}{\omega x_{i}^2 \over 2cf}\left[{d_{2}\over f}-1\right]}e^{{\rm
i}{\omega x^2 \over 2cd_{1}}}\int dx' p(x')e^{{\rm i}{\omega x'^2
\over 2cd_{1}}}e^{{\rm -i}{{\omega \over c}x'\left[{x\over
d_{1}}+{x_{i}\over f}\right]}}\,,
\end{equation}

\noindent which is the impulse response function of the entire
optical system from the crystal output plane to the detector input
plane. We use this impulse response function to determine the
transfer function of the system in terms of transverse wavevectors
via

\begin{equation}
\label{Transfer-Function-Definition} H({\bf x}_{i},{\bf
q};\omega)=\int d{\bf x} \,h({\bf x}_{i},{\bf x};\omega)\,e^{{\rm
i}{\bf q}\cdot{\bf x}}\,,
\end{equation}

\noindent so that the transfer function explicitly takes the form

\begin{equation}
\label{Transfer-Function-Explicit} H({\bf x}_{i},{\bf
q};\omega)=\left[e^{{\rm i}{\omega \over c}(d_{1}+d_{2}+f)}e^{{\rm
-i}{\omega |{\bf x}_{i}|^2 \over 2cf}\left[{d_{2}\over
f}-1\right]}e^{{\rm -i}{cd_{1} \over 2\omega}|{\bf
q}|^2}\tilde{P}\left({\omega \over cf}{\bf x}_{i}  -{\bf
q}\right)\right]\,{\cal F}(\omega)\,,
\end{equation}

\noindent where the function $\tilde{P}\left({\omega \over cf}{\bf
x}_{i} -{\bf q}\right)$ is defined by

\begin{equation}
\label{Aperture-Function-Definition} \tilde{P}\left({\omega \over
cf}{\bf x}_{i}  -{\bf q}\right)=\int d{\bf x'} p({\bf x'})e^{{\rm
-i}{\omega {\bf x'}\cdot{\bf x}_{i}\over cf}}e^{{\rm i}{\bf
q}\cdot{\bf x'}}\,.
\end{equation}

\noindent Using Eq.~(\ref{Transfer-Function-Explicit}) we can now
describe the propagation of the down-converted light from the
crystal to the detection planes. Since no birefringence is assumed
for any material in the system considered to this point, this
transfer function is identical for both polarization modes ({\em
o},{\em e}). In some of the experimental arrangements discussed in
this paper, a prism is used to separate the pump field from the
SPDC. The alteration of the transfer function $H$ by the presence
of this prism is found mathematically to be negligible (See
Appendix~A) and the effect of the prism is neglected. A parallel
set of experiments conducted without the use of a prism further
justifies this conclusion [See section III.(C)].

Continuing the analysis in the Fresnel approximation and the
approximation that the SPDC fields are quasi-monochromatic, we can
derive an analytical form for the coincidence-count rate defined
in Eq.~(\ref{Coincidence-Slow}):

\begin{equation}
\label{Coincidence-filter}
 R (\tau)= R_{0}\left[1-V\left(\tau\right)\right]\,,
\end{equation}

\noindent where $R_{0}$ is the coincidence rate outside the region
of quantum interference. In the absence of spectral filtering

\begin{equation}\label{V-transverse}
    V(\tau)=\Lambda \left(\frac{2\tau}{LD}-1\right)\
                  {\rm sinc}\left[\frac{\omega_{p}^{0}L^{2} M^{2}}{4cd_{1}}\frac{\tau}{LD}\
                        \Lambda \left(\frac{2\tau}{LD}-1\right)\right]\
                  \tilde{{\cal P}}_{A}\left(-\frac{\omega_{p}^{0}LM}{4cd_{1}}\frac{2\tau}{LD}{\bf e}_{2}\right)\
                  \tilde{{\cal P}}_{B}\left(\frac{\omega_{p}^{0}LM}{4cd_{1}}\frac{2\tau}{LD}{\bf
                  e}_{2}\right)\,,
\end{equation}

\noindent where $D=1/u_{o}-1/u_{e}$ with $u_{j}$ denoting the
group velocity for the $j$-polarized photon ($j=o,e$), $M=\partial
\ln n_{e}(\omega^{0}_p / 2,\theta_{\rm OA})/{\partial \theta_{e}}$
\cite{M-Parameter}, and $\Lambda (x)=1-|x|$ for $-1\leq x \leq 1$,
and zero otherwise. A derivation of Eq.~(\ref{V-transverse}),
along with the definitions of all quantities in this expression,
is presented in Appendix B. The function ${\tilde{\cal P}}_{i}$
(with $i=A,B$) is the normalized Fourier transform of the squared
magnitude of the aperture function $p_{i}({\bf x})$; it is given
by

\begin{equation}\label{P-d}
  \tilde{{\cal P}}_{i}({\mathbf q})=\frac{
        \int \int d{\mathbf y}\ p_{i}({\mathbf y})p^{\ast}_{i}({\mathbf y})\
            e^{-{\rm i}{\mathbf y}\cdot{\mathbf q}}}{
        \int \int d{\mathbf y}\ p_{i}({\mathbf y})p^{\ast}_{i}({\mathbf
        y})}\,.
\end{equation}

The profile of the function ${\tilde{\cal P}}_{i}$ within Eq.
(\ref{V-transverse}) plays a key role in the results presented in
this paper. The common experimental practice is to use extremely
small apertures to reach the one-dimensional planewave limit. As
shown in Eq.~(\ref{V-transverse}), this gives ${\tilde{\cal
P}}_{i}$ functions that are broad in comparison with $\Lambda$ so
that $\Lambda$ determines the shape of the quantum interference
pattern, resulting in a symmetric triangular dip. The sinc
function in Eq.~(\ref{V-transverse}) is approximately equal to
unity for all practical experimental configurations and therefore
plays an insignificant role. On the other hand, this sinc function
represents the difference between the familiar one-dimensional
model (which predicts $R(\tau) = R_0 \left[1-\Lambda
\left(\frac{2\tau}{LD}-1\right)\right]$, a perfectly triangular
interference dip) and a three-dimensional model in the presence of
a very small on-axis aperture.

Note that for symmetric apertures $|p_{i}({\bf x})|=|p_{i}(-{\bf
x})|$, so from Eq.~(\ref{P-d}) the functions $\tilde{{\cal
P}}_{i}$ are symmetric as well. However, within
Eq.~(\ref{V-transverse}) the $\tilde{{\cal P}}_{i}$ functions,
which are centered at $\tau=0$, are shifted with respect to the
function $\Lambda$, which is symmetric about $\tau= LD/2$. Since
Eq.~(\ref{V-transverse}) is the product of functions with
different centers of symmetry, it predicts asymmetric
quantum-interference patterns, as have been observed in recent
experiments \cite{Femto-PRL}. When the apertures are wide, the
$p_{i}({\bf x})$ are broad functions which result in narrow
${\tilde{\cal P}}_{i}$, so that the interference pattern is
strongly influenced by the shape of the functions ${\tilde{\cal
P}}_{i}$. If, in addition, the apertures are spatially shifted in
the transverse plane, the ${\tilde{\cal P}}_{i}$ become
oscillatory functions that result in sinusoidal modulation of the
interference pattern. This can result in a partial inversion of
the dip into a peak for certain ranges of the delay $\tau$, as
will be discussed subsequently. In short, it is clear from
Eq.~(\ref{V-transverse}) that $V(\tau)$ can be altered
dramatically by carefully selecting the aperture profile.

When SPDC is generated using a finite-bandwidth pulsed pump field,
Eq. (\ref{V-transverse}) becomes

\begin{equation}
\label{V-transverse-pulse}    V(\tau)=\Lambda
\left(\frac{2\tau}{LD}-1\right)\
                  {\cal V}_{p}(\tau)~\tilde{{\cal P}}_{A}\left(-\frac{\omega_{p}^{0}LM}{4cd_{1}}\frac{2\tau}{LD}{\bf e}_{2}\right)\
                  \tilde{{\cal P}}_{B}\left(\frac{\omega_{p}^{0}LM}{4cd_{1}}\frac{2\tau}{LD}{\bf
                  e}_{2}\right)\,,
\end{equation}

\noindent where the sinc function in Eq. (\ref{V-transverse}) is
simply replaced by ${\cal V}_{p}$, which is given by

\begin{equation}
\label{Vp-pulse} {\cal V}_{p}(\tau)=\frac{\int d\omega_{p}~|{\cal
E}_{p}(\omega_{p}-\omega_{p}^{0})|^2{\rm
sinc}~\left[[D_{+}L(\omega_{p}-\omega_{p}^{0}) +
\frac{\omega_{p}^{0}L^{2} M^{2}}{4cd_{1}}\frac{\tau}{LD}] \Lambda
\left(\frac{2\tau}{LD}-1\right)\right]}{\int d\omega_{p}~|{\cal
E}_{p}(\omega_{p}-\omega_{p}^{0})|^2}
\end{equation}

\noindent with $D_{+}=1/u_{p}-\frac{1}{2}(1/u_{o}+1/u_{e})$ where
$u_{j}$ denotes the group velocity for the $j$-polarized photon
($j=p,~o,{\rm ~and~} e$) and all other parameters are identical to
those in Eq.~(\ref{V-transverse}). The visibility of the dip in
this case is governed by the bandwidth of the pump field.

\subsection{Quantum Interference with Circular Apertures}

Of practical interest are the effects of the aperture shape and
size, via the function $\tilde{P}({\bf q})$, on polarization
quantum-interference patterns. To this end, we consider the
experimental arrangement illustrated in Fig.~3(a) in the presence
of a circular aperture with diameter $b$. The mathematical
representation of this aperture is given in terms of the Bessel
function $J_{1}$,

\begin{equation}
\label{Aperture-Function-Circular} \tilde{{\cal P}}({\bf
q})=2\,{J_{1}\left(b\left|{\bf q}\right|\right) \over b\left|{\bf
q}\right|}\,.
\end{equation}

For the experiments conducted with cw-pumped SPDC the pump was a
single-mode cw argon-ion laser with a wavelength of 351.1~nm and a
power of 200~mW. The pump light was delivered to a $\beta$-${\rm
BaB_2O_4}$ (BBO) crystal with a thickness of 1.5~mm. The crystal
was aligned to produce collinear and degenerate Type-II
spontaneous parametric down-conversion. Residual pump light was
removed from the signal and idler beams with a fused-silica
dispersion prism. The collinear beams were then sent through a
delay line comprised of a z-cut crystalline quartz element (fast
axis orthogonal to the fast axis of the BBO crystal) whose
thickness was varied to control the relative optical-path delay
between the photons of a down-converted pair. The characteristic
thickness of the quartz element was much less then the distance
between the crystal and the detection planes, yielding negligible
effects on the spatial properties of the SPDC. The photon pairs
were then sent to a non-polarizing beam splitter. Each arm of the
polarization intensity interferometer following this beam splitter
contained a Glan-Thompson polarization analyzer set to $45^\circ$,
a convex lens to focus the incoming beam, and an actively quenched
Peltier-cooled single-photon-counting avalanche photodiode
detector [denoted ${\rm D}_i$ with $i=A,B$ in Fig.~3(a)]. No
spectral filtering was used in the selection of the signal and
idler photons for detection. The counts from the detectors were
conveyed to a coincidence counting circuit with a 3-ns
coincidence-time window. Corrections for accidental coincidences
were not necessary.

The experiments with pulse-pumped SPDC were carried out using the
same interferometer as that used in the cw-pumped SPDC
experiments, but with ultrafast laser pulses in place of cw laser
light. The pump field was obtained by frequency doubling the
radiation from an actively mode-locked Ti:Sapphire laser, which
emitted pulses of light at 830 nm. After doubling, 80-fsec pulses
(FWHM) were produced at 415 nm, with a repetition rate of 80 MHz
and an average power of 15 mW.

The observed normalized coincidence rates (quantum-interference
patterns) from a cw-pumped 1.5-mm BBO crystal (symbols), along
with the expected theoretical curves (solid), are displayed in
Fig.~4 as a function of relative optical-path delay for various
values of aperture diameter $b$ placed 1~m from the crystal.
Clearly the observed interference pattern is more strongly
asymmetric for larger values of $b$. As the aperture becomes
wider, the phase-matching condition between the pump and the
generated down-conversion allows a greater range of $({\bf
q},\omega)$ modes to be admitted. The $({\bf q},\omega)$ modes
that overlap less introduce more distinguishability. This inherent
distinguishability, in turn, reduces the visibility of the
quantum-interference pattern and introduces an asymmetric shape.

The theoretical plots of the visibility of the
quantum-interference pattern at the full-compensation delay
$\tau=LD/2$, as a function of the crystal thickness, is plotted in
Fig.~5 for various aperture diameters placed 1~m from the crystal.
Full visibility is expected only in the limit of extremely thin
crystals, or with the use of an extremely small aperture where the
one-dimensional limit is applicable. As the crystal thickness
increases, the visibility depends more dramatically on the
aperture diameter.  The experimentally observed visibility for
various aperture diameters, using the 1.5-mm thick BBO crystal
employed in our experiments (symbols), agrees well with the
theory.

If the pump field is pulsed, then there are additional limitations
on the visibility that emerge as a result of the broad spectral
bandwidth of the pump field \cite{Femto-PRL,Femto-SPDC,Perina}.
The observed normalized coincidence rates (symbols) from a pump of
80-fsec pulses and a 5-mm aperture placed 1 m from the crystal,
along with the expected theoretical curves (solid), assuming a
Gaussian spectral profile for the pump, are displayed in Fig.~6 as
a function of relative optical-path delay for 0.5-mm, 1.5-mm, and
3-mm thick BBO crystals. The asymmetry of the interference pattern
for increasing crystal thickness is even more visible in the
pulsed regime. Figure 7 shows a visibility plot similar to Fig.~5
for the pulsed-pump case.

\subsection{Quantum Interference with Slit Apertures}

For the majority of quantum-interference experiments involving
relative optical-path delay, circular apertures are the norm. In
this section we consider the use a vertical slit aperture to
investigate the transverse symmetry of the generated photon pairs.
Since the experimental arrangement of Fig.~3(a) remains identical
aside from the aperture, Eq.~(\ref{Biphoton-Separable-1}) still
holds and $p({\bf x})$ takes the explicit form

\begin{equation}
\label{Aperture-Function-Vertical-Slit} \tilde{{\cal P}}({\bf
q})={\sin(b\, {\bf e}_{1} \cdot {\bf q}) \over b\, {\bf e}_{1}
\cdot {\bf q}}\,{\sin(a \,{\bf e}_{2} \cdot {\bf q}) \over a\,
{\bf e}_{2} \cdot {\bf q}}\,.
\end{equation}

\noindent The data shown by squares in Fig.~8 is the observed
normalized coincidence rate for a cw-pumped 1.5-mm BBO in the
presence of a vertical slit aperture with $a=7$ mm and $b=1$ mm.
The quantum-interference pattern is highly asymmetric and has low
visibility, and indeed is similar to that obtained using a wide
circular aperture (see Fig.~4). The solid curve is the theoretical
quantum-interference pattern expected for the vertical slit
aperture used.

In order to investigate the transverse symmetry, the complementary
experiment has also been performed using a horizontal slit
aperture. For the horizontal slit, the parameters $a$ and $b$ in
Eq.~(\ref{Aperture-Function-Vertical-Slit}) are interchanged so
that $a=1$ mm and $b=7$ mm. The data shown by triangles in Fig.~8
is the observed normalized coincidence rate for a cw-pumped 1.5-mm
BBO in the presence of this aperture. The most dramatic effect
observed is the symmetrization of the quantum-interference pattern
and the recovery of the high visibility, despite the wide aperture
along the horizontal axis. A practical benefit of such a slit
aperture is that the count rate is increased dramatically, which
is achieved by limiting the range of transverse wavevectors along
the optical axis of the crystal to induce indistinguishability and
allowing a wider range along the orthogonal axis to increase the
collection efficiency of the SPDC photon pairs. This finding is of
significant value, since a high count rate is required for many
applications of entangled photon pairs and, indeed, many
researchers have suggested more complex means of generating
high-flux photon pairs \cite{New-Source}.

Noting that the optical axis of the crystal falls along the
vertical axis, these results verify that the dominating portion of
distinguishability lies, as expected, along the optical axis. The
orthogonal axis (horizontal in this case) provides a negligible
contribution to distinguishability, so that almost full visibility
can be achieved despite the wide aperture along the horizontal
axis.

The optical axis of the crystal in the experimental arrangements
discussed above is vertical with respect to the lab frame. This
coincides with the polarization basis of the down-converted
photons. To stress the independence of these two axes, we wish to
make the symmetry axis for the spatial distribution of SPDC
distinct from the polarization axes of the down-converted photons.
One way of achieving this experimentally is illustrated in
Fig.~9(a). Rather than using a simple BBO crystal we incorporate
two half-wave plates, one placed before the crystal and aligned at
$22.5^\circ$ with respect to the vertical axis of the laboratory
frame, and the other after the crystal and aligned at
$-22.5^\circ$. The BBO crystal is rotated by $45^\circ$ with
respect to the vertical axis of the laboratory frame.
Consequently, SPDC is generated in a special distribution with a
$45^\circ$-rotated axis of symmetry, while keeping the
polarization of the photons aligned with the horizontal and the
vertical axes. Rotated-slit-aperture experiments with cw-pumped
SPDC, similar to those presented in Fig.~8, were carried out using
this arrangement. The results are shown in Fig.~9(b). The highest
visibility in the quantum-interference pattern occurred when the
vertical slit was rotated $-45^\circ$ (triangles), while the
lowest and most asymmetric pattern occurred when the vertical slit
aperture was rotated $+45^\circ$ (squares). This verifies that the
effect of axis selection is solely due to the spatiotemporal
distribution of SPDC, and not related to any birefringence or
dispersion effects associated with the linear elements in the
experimental arrangement.

\subsection{Quantum-Interference with Increased Acceptance Angle}

A potential obstacle for accessing a wider range of transverse
wavevectors is the presence of dispersive elements in the optical
system. One or more dispersion prisms, for example, are often used
to separate the intense pump field from the down-converted photons
\cite{Dispersion-Book}. As discussed in Appendix A, the finite
angular resolution of the system aperture and collection optics
can, in certain limits, cause the prism to act as a spectral
filter.

To increase the limited acceptance angle of the detection system
and more fully probe the multi-parameter interference features of
the entangled-photon pairs, we carry out experiments using the
alternate setup shown in Fig.~10. Note that a dichroic mirror is
used in place of a prism. Moreover, the effective acceptance angle
is increased by reducing the distance between the crystal and the
aperture plane. This allows us to access a greater range of
transverse wavevectors with our interferometer, facilitating the
observation of the effects discussed in sections III.(A) and
III.(B) without the use of a prism.

Using this experimental arrangement, we repeated the
circular-aperture experiments, the results of which were presented
in Fig.~4. Figure~11 displays the observed quantum-interference
patterns (normalized coincidence rates) from a cw-pumped 1.5-mm
BBO crystal (symbols) along with the expected theoretical curves
(solid) as a function of relative optical-path delay for various
values of the aperture diameter $b$ placed 750~mm from the
crystal. For the data on the curve with the lowest visibility
(squares), the limiting apertures in the system were determined
not by the irises as shown in Fig.~11, but by the dimensions of
the Glan-Thompson polarization analyzers, which measure 7~mm
across.

Similar experiments were conducted with pulse-pumped SPDC in the
absence of prisms. The resulting quantum-interference patterns are
shown as symbols in Fig.~12. The solid curves are the interference
patterns calculated by using the model given in Section II, again
assuming a Gaussian spectral profile for the pump. Comparing
Fig.~12 with Fig.~6, we note that the asymmetry in the
quantum-interference patterns is maintained in the absence of a
prism, verifying that this effect is consistent with the
multi-parameter entangled nature of SPDC.

An experimental study directed specifically toward examining the
role of the prism in a similar quantum-interference experiment has
recently been presented \cite{Dispersion-Experiment}. In that
work, the authors carried out a set of experiments both with and
without prisms. They reported that the interference patterns
observed with a prism in the apparatus are asymmetric, while those
obtained in the absence of such a prism are symmetric. The authors
claim that the asymmetry of the quantum-interference pattern is an
artifact of the presence of the prism. In contrast to the
conclusions of that study, we show that asymmetrical patterns are,
in fact, observed in the absence of a prism [see Fig.~(12)].
Indeed, our theory and experiments show that interference patterns
become symmetric when narrow apertures are used, either in the
absence or in the presence of a prism. This indicates conclusively
that transverse effects alone are responsible for asymmetry in the
interference pattern. In the experiment reported in
Ref.~\cite{Dispersion-Experiment}, the coincidence-count rate
underwent an unexplained fiftyfold reduction (from 126~counts/sec
down to 2.4~counts/sec) as the pattern returned to symmetric form
despite the fact that a nonabsorbing prism was used. This decrease
in the count rate suggests that considerably different acceptance
angles were used in these experiments and the results were
improperly ascribed to the presence of the prism.

\subsection{Pump-Field Diameter Effects}

The examples of the pump field we have considered are all
planewaves. In this section, and in the latter part of Appendix B,
we demonstrate the validity of this assumption under our
experimental conditions and find a limit where this assumption is
no longer valid. To demonstrate the independence of the
interference pattern on the size of the pump, we placed a variety
of apertures directly at the front surface of the crystal.
Figure~13 shows the observed normalized coincidence rates from a
cw-pumped 1.5-mm BBO crystal (symbols) as a function of relative
optical-path delay for various values of pump beam diameter. The
acceptance angle of the optical system for the down-converted
light is determined by a 2.5-mm aperture at a distance of 750~mm
from the crystal. The theoretical curve (solid) corresponds to the
quantum-interference pattern for an infinite planewave pump.
Figure~14 shows similar plots as in Fig.~13 in the presence of a
5-mm aperture in the optical system for the down-converted light.
The typical value of the pump beam diameter in
quantum-interference experiments is 5~mm. The experimental results
from a 5-mm, 1.5-mm, and 0.2-mm diameter pump all lie within
experimental uncertainty, and are practically identical aside from
the extreme reduction in count rate due to the reduced pump
intensity.

This behavior of the interference pattern suggests that the
dependence of the quantum-interference pattern on the diameter of
the pump beam is negligible within the limits considered in this
work. Indeed, if the pump diameter is comparable to the spatial
walk-off of the pump beam within the nonlinear crystal, then the
planewave approximation is not valid and the proper spatial
profile of the pump beam must be considered in
Eq.~(\ref{Biphoton-Separable-Spatial}). For the 1.5-mm BBO used in
our experiments this limit is approximately 70 $\mu$m, which is
smaller than any aperture we could use without facing
prohibitively low count rates.

\subsection{Shifted-Aperture Effects}

In the work presented thus far, the optical elements in the system
are placed concentrically about the longitudinal ({\em z}) axis.
In this condition, the sole aperture before the beam splitter, as
shown in Fig.~3(a), yields the same transfer function as two
identical apertures placed in each arm after the beam splitter, as
shown in Fig.~10. In this section we show that the observed
quantum-interference pattern is also sensitive to a {\em relative
shift} of the apertures in the transverse plane. To account for
this, we must include an additional factor in
Eq.~(\ref{V-transverse}):

\begin{equation}\label{cos-sa-sb}
  \cos\left[\frac{\omega_{p}LM}{4cd_{1}}\frac{2\tau}{LD}{\bf e}_{2} \cdot ({\bf s }_{A}-{\bf
  s}_{B})\right]\,,
\end{equation}
where ${\bf s }_{i}$ (with $i=A,B$) is the displacement of each
aperture from the longitudinal ({\em z}) axis. This extra factor
provides yet another degree of control on the quantum-interference
pattern for a given aperture form.

\subsubsection{Quantum Interference with Shifted-Slit Apertures}

First, we revisit the case of slit apertures, discussed above in
Section III.(B). Using the setup shown in Fig.~10, we placed
identically oriented slit apertures in each arm of the
interferometer, which can be physically shifted up and down in the
transverse plane. A spatially shifted aperture introduces an extra
phase into the $\tilde{{\cal P}}({\bf q})$ functions, which in
turn results in the sinusoidal modulation of the
quantum-interference pattern as shown in Eq.~(\ref{cos-sa-sb})
above.

The two sets of data shown in Fig.~15 represent the observed
normalized coincidence rates for a cw-pumped 1.5-mm BBO crystal in
the presence of identical apertures placed without shift in each
arm as shown in Fig.~10. The triangular points correspond to the
use of 1$\times$7~mm horizontal slits. The square points
correspond to the same apertures rotated 90 degrees to form
vertical slits. Since this configuration, as shown in Fig.~10,
accesses a wider range of acceptance angles, the dimensions of the
other optical elements become relevant as effective apertures in
the system. Although the apertures themselves are aligned
symmetrically, an effective vertical shift of $\left|{\bf
s}_A-{\bf s}_B\right|=1.6$~mm is induced by the relative
displacement of the two polarization analyzers. The solid curves
in Fig.~15 are the theoretical plots for the two aperture
orientations. Note that as in the experiments described in Section
III.(B), the horizontal slits give a high visibility interference
pattern, and the vertical slits give an asymmetric pattern with
low visibility, even in the absence of a prism. Note further that
the cosine modulation of Eq.~(\ref{cos-sa-sb}) results in peaking
of the interference pattern when the vertical slits are used.

\subsubsection{Quantum Interference with Shifted-Ring Apertures}

Given the experimental setup shown in Fig.~10 with an annular
aperture in arm A and a 7~mm circular aperture in arm B, we
obtained the quantum-interference patterns shown in Fig.~16. The
annular aperture used had an outer diameter of $b=4$~mm and an
inner diameter of $a=2$~mm, yielding an aperture function

\begin{equation}
\label{Aperture-Function-Ring-2}
 \tilde{{\cal P}}_{A}({\bf q})={2 \over {b-a}}\,\left[\frac{J_{1}\left(b\left|{\bf
 q}\right|\right)}{\left|{\bf q}\right|}-\frac{J_{1}\left(a\left|{\bf
 q}\right|\right)}{\left|{\bf q}\right|}\right].
\end{equation}

\noindent The symbols give the experimental results for various
values of the relative shift $\left|{\bf s}_A-{\bf s}_B\right|$,
as denoted in the legend.  Note that as in the case of the shifted
slit, $V(\tau)$ becomes negative for certain values of the
relative optical-path delay ($\tau$), and the interference pattern
displays a peak rather than the familiar triangular dip usually
expected in this type of experiment.

\section{Conclusion}

In summary, we observe that the multi-parameter entangled nature
of the two-photon state generated by SPDC allows transverse
spatial effects to play a role in polarization-based quantum
interference experiments.  The interference patterns generated in
these experiments are, as a result, governed by the profiles of
the apertures in the optical system which admit wavevectors in
specified directions. Including a finite bandwidth for the pump
field strengthens this dependence on the aperture profiles,
clarifying why the asymmetry was first observed in the ultrafast
regime \cite{Femto-PRL}. The effect of the pump-beam diameter on
the quantum-interference pattern is shown to be negligible for a
typical range of pump diameter values used in similar experimental
arrangements in the field. The quantitative agreement between the
experimental results using a variety of aperture profiles, and the
theoretical results from the formalism presented in Section II of
this paper confirm this interplay.  In contrast to the usual
single-direction polarization entangled state, the wide-angle
polarization entangled state offers a richness that can be
exploited in a variety of applications involving quantum
information processing.

\appendix

\section*{(A) Effect of Prism on System Transfer Function}

We present a mathematical analysis of the effect of the prism used
in some of the experiments presented above on the spatiotemporal
distribution of SPDC. We begin by assuming that the central
wavelength for down-conversion is aligned at the minimum deviation
angle $\phi_0$, so that the input and the output angles of the
prism are equal. The optical axis ($z$) used in the calculation of
the system transfer function follows these angles as shown in the
inset of Fig.~17. Within the paraxial and quasi-monochromatic
field approximations the prism is represented by a mapping of each
$({\bf q},\omega)$ mode to a $({\bf q'},\omega)$ mode. Using
Snell's Law, the relation between ${\bf q}$ and ${\bf q'}$ at a
given frequency $\omega$ is dictated by

\begin{equation}
\label{Prism-Mapping} {\cal S}({\bf q},{\bf q'}) + {\cal S}({\bf
q'},{\bf q}) = {n}^2(\omega) \sin(\alpha)\,,
\end{equation}

\noindent where $\alpha$ is the apex angle of the prism. The
function ${\cal S}$ is given by

\begin{eqnarray}
{\cal S}({\bf q},{\bf
q'})=&&\left[\sin(\phi_{0})+{\cos(\phi_{0})\,c \over \omega} {\bf
q}\cdot{\bf e}_{1}-{\sin(\phi_{0})c^2\over 2\omega^2}|{\bf
q}\cdot{\bf e}_{1}|^2\right]
\nonumber \\
&&\times \left[ {n}^2(\omega) - {\sin(\phi_{0})}^2 -
{\sin(2\phi_{0})\,c \over \omega}{\bf q'}\cdot{\bf e}_{1}
-{\cos(2\phi_{0})c^2 \over \omega^2}|{\bf q'}\cdot{\bf
e}_{1}|^2\right]^{1 \over 2}\,.
\end{eqnarray}

Figure 17 shows a plot of ${\bf q'}$ as a function of ${\bf q}$.
Plotting this curve for various frequencies $\omega$ produces no
deviation visible within the resolution of the printed graph. The
small box at the center of the plot highlights the range of
transverse wavevectors limited by the acceptance angle of the
optical system used in the above experiments. The dashed line is
the $|{\bf q'}|=|{\bf q}|$ curve with unity slope.

We now consider a linear expansion in frequencies and transverse
wavevectors of the above-mentioned mapping of a $({\bf q},\omega)$
mode to a $({\bf q'},\omega)$ mode in the form

\begin{eqnarray}
\label{linear-prism} {\bf q'}\cdot{\bf e}_1 & \approx & -{\bf
q}\cdot{\bf e}_1 +\beta \, \frac{\omega_{p}^0}{c} \left(\omega-{\omega_{p}^0 \over 2}\right) \nonumber \\
{\bf q'}\cdot{\bf e}_2 & \approx & {\bf q}\cdot{\bf e}_2\,,
\end{eqnarray}

\noindent where the negative sign multiplying ${\bf q}\cdot{\bf
e}_1$ indicates that a ray of light at the input face of the prism
with a small deviation in one direction is mapped to a ray of
light at the output face with a corresponding deviation in the
opposite direction [See Eq.~(16) in Section 4.7.2 of
Ref.~\cite{Born-Wolf}]. The parameter $\beta$ corresponds to the
angular dispersion parameter of the prism \cite{Born-Wolf} with
the explicit form given by

\begin{equation}
\label{prism-beta} \beta={{\sin \alpha} \over {\cos \phi_{0} \,
\cos(\alpha/2)}} \,{dn \over d\omega}\,.
\end{equation}

To find the range of aperture diameters $b$ where the effect of
the prism can be considered negligible, we need to compare the
angular resolution of the aperture-lens combination at the
detection plane to the angular dispersion of the prism. A system
with an infinite aperture and an infinite lens maps each
wavevector into a distinct point at the detection plane. In
practice, of course, a finite aperture limits the angular
resolution of the system at the detection plane to the order of
$2\lambda/b$, where $\lambda$ is the central wavelength of the
down-converted light and $b$ is the diameter of the aperture. Now,
the prism maps each frequency into a distinct wavevector. The
angular dispersion of the prism is on the order of
$\beta\,\Delta\omega$ where $\Delta\omega$ is the bandwidth of the
incident light beam. If the angular dispersion of the prism is
less than the angular resolution of the combined aperture and
lens, the dispersive properties of the prism have negligible
effect on the quantum interference pattern.

For an SPDC bandwidth of 10~nm around a central wavelength of
702~nm and a value of $\beta$ given by $5.8\times10^{-18}$~sec
(calculated from the material properties of fused silica), the
effect of the angular dispersion introduced by the prism on the
experiments presented in this paper can be safely neglected for
aperture diameters less than 20~mm. For aperture diameters in the
vicinity of this value and higher, the effect on the
quantum-interference pattern is a spectral-filter-like smoothing
of the edges. The chromatic dispersion experienced by the
down-converted light when propagating through such dispersive
elements was previously examined~\cite{Perina}; the dispersiveness
of the material must be an order of magnitude higher than the
values used in the experiments considered here to have a
significant effect on the down-converted light. Pursuant to these
arguments, the transfer function of the system, as provided in
Eq.~(\ref{Transfer-Function-Explicit}), is not affected by the
presence of the prism. Moreover, the results discussed in sections
III.(A), III.(B), and III.(C) experimentally confirm that the
effect of the prism on the quantum-interference pattern is
negligible in comparison with other spatial and spectral effects.

\section*{(B) Derivation of Visibility in Eq.~(32)}

The purpose of this Appendix is to derive Eq.~(\ref{V-transverse})
using Eqs.~(\ref{Coincidence-Slow}),~(\ref{Biphoton-Separable-1}),
and~(\ref{Transfer-Function-Explicit}). To obtain an analytical
solution within the Fresnel approximation we assume
quasi-monochromatic fields and perform an expansion in terms of a
small angular frequency spread ($\nu$) around the central angular
frequency ($\omega^{0}_{p}/2$) associated with degenerate
down-conversion, and small transverse components $|{\mathbf q}|$
with respect to the total wavevector ${\bf k}_{j}$ for collinear
down-conversion. In short, we use the fact that $|\nu| \ll
\omega^{0}_{p}/2$, with $\nu=\omega-\omega^{0}_{p}/2$, and
$|{\mathbf q}|^{2}\ll |{\mathbf k}|^{2}$. In these limits we
obtain
\begin{eqnarray}
\label{kz-appr} \kappa_{o}\left(\omega, {\mathbf q}\right)
        &&\approx
        K_{o}+\frac{\omega-\omega^{0}_{p}/2}{u_{o}}-\frac{|{\mathbf q}|^{2}}{2K_{o}} \\
\label{kz-appr2}
        \kappa_{e}\left(\omega, {\mathbf q}\right)
        && \approx
        K_{e}-\frac{\omega-\omega^{0}_{p}/2}{u_{e}}-\frac{|{\mathbf q}|^{2}}{2K_{e}}+M{\mathbf e}_{2}\cdot{\mathbf
        q}\,,
\end{eqnarray}
where the explicit forms for $K_{j}$, $u_{j}$ $j = o,e$, and
$M{\mathbf e}_{2}$ are \cite{M-Parameter}:
\begin{eqnarray}\label{Param}
    K_{j}&=&\left. |{\bf k}_{j}|(\omega, {\mathbf q})\right|_{{\omega^{0}_{p}\over 2},{\mathbf q}=0}
        \hspace{2.9cm} \frac{1}{u_{j}}=\left.\frac{\partial  \kappa_{j}(\omega, {\mathbf q})}
                                {\partial \omega}\right|_{{\omega^{0}_{p} \over 2},{\mathbf q}=0} \nonumber \\
    M {\mathbf e}_{2}&=&\left. \frac{|{\bf k}_{e}| \nabla_{{\mathbf q}}|{\bf k}_{e}|}{\kappa_{e}}\right|_{{\omega^{0}_p \over 2},{\mathbf q}=0}
    \hspace{1.75cm} M=\left. \frac{\partial \ln
        n_{e}(\omega,\theta_{e})}{\partial
        \theta_{e}}\right|_{{\omega^{0}_p \over 2}, \theta_{e}=\theta_{\rm OA}}\,.
\end{eqnarray}

\noindent Using the results in
Eqs.~(\ref{kz-appr}),~(\ref{kz-appr2}) and (\ref{Param}) we can
now provide an approximate form for $\Delta$, which is the
argument of the sinc function in Eq. (\ref{Biphoton-Separable-1}),
as

\begin{equation}\label{Delta-appr}
  \Delta \approx
    - D\nu +
    \frac{2c|{\mathbf q}|^{2}}{\omega_{p}^{0}}+
    M{\mathbf e}_{2}\cdot{\mathbf q}\,,
\end{equation}

\noindent where $D=\frac{1}{u_{o}}-\frac{1}{u_{e}}$.

Using this approximate form for $\Delta$ in the integral
representation of ${\rm sinc}(x)$

\begin{equation}\label{sinc-function}
  {\rm sinc}\left({{L\Delta}\over2}\right)
            e^{{\rm -i}{{L\Delta}\over2}}=
          \int^{0}_{-L} dz ~e^{-{\rm i}z\Delta}\,,
\end{equation}

\noindent with the assumption that $L\ll d_{1}$, we obtain
Eq.~(\ref{Coincidence-filter}) with

\begin{eqnarray}\label{R-inf-new}
   R_{0}=\int d\nu \int_{-L}^{0} dz\
 e^{-{\rm i}D\nu z}\int_{-L}^{0} dz^{\prime}\
 e^{{\rm i}D\nu z^{\prime}}\
 {\mathcal J}_{0}(z,z^{\prime})\,,
\end{eqnarray}

\begin{eqnarray}\label{R-0-new}
  V(\tau)={1 \over  R_{0}}\,\int d\nu\
 e^{-2i\tau\nu} \int_{-L}^{0} dz\
 e^{-{\rm i}D\nu z}\int_{-L}^{0} dz^{\prime}\
 e^{-{\rm i}D\nu z^{\prime}}\
  {\mathcal J}_{V}(z,z^{\prime})\,,
\end{eqnarray}

\noindent where the functions

\begin{eqnarray}\label{G-inf-appr}
   {\mathcal J}_{0}(z,z^{\prime})&=&
          \left(\frac{\omega^{0}_{p}}{2cd_{1}}\right)^{2}\
          \exp \left[-{\rm i}\frac{\omega^{0}_{p}}{8cd_{1}}M^{2}(z^{2}-
          z^{\prime\ 2}) \right]\nonumber\\ &&
    \times
         \tilde{{\cal P}}_{A}\left[\frac{\omega^{0}_{p}}{4cd_{1}}M(z- z^{\prime}){\bf e}_{2}\right]\
          \tilde{{\cal P}}_{B}\left[-\frac{\omega^{0}_{p}}{4cd_{1}}M(z-
          z^{\prime}){\bf e}_{2}\right]\,,
\end{eqnarray}

\begin{eqnarray}\label{G-0-appr}
   {\mathcal J}_{V}(z,z^{\prime})&=&
          \left(\frac{\omega^{0}_{p}}{2cd_{1}}\right)^{2}\
          \exp \left[-{\rm i}\frac{\omega^{0}_{p}}{8cd_{1}}M^{2}(z^{2}-
          z^{\prime\ 2}) \right]\nonumber\\ &&
    \times
         \tilde{{\cal P}}_{A}\left[\frac{\omega^{0}_{p}}{4cd_{1}}M(z+ z^{\prime}){\bf e}_{2}\right]\
          \tilde{{\cal P}}_{B}\left[-\frac{\omega^{0}_{p}}{4cd_{1}}M(z+
          z^{\prime}){\bf e}_{2}\right]\,,
\end{eqnarray}

\noindent are derived by carrying out the integrations over the
variables ${\bf x}$ and ${\bf q}$.

Performing the remaining integrations leads us to Eq.
(\ref{V-transverse}). This equation allows robust and rapid
numerical simulations of quantum-interferometric measurement to be
obtained. For the simulations provided in this work with pump
wavelength of 351~nm, the calculated values of $M$ and $D$ are
0.0711 and 248 fsec/mm, respectively. Similarly, given a pump
wavelength of 415~nm, we compute $M$ = 0.0723 and $D$ = 182
fsec/mm.

In the case of type-I SPDC, both photons of a generated pair have
ordinary polarization. Consequently, the vector $M{\bf e}_{2}$
does not appear in the expansions of the wavevectors, unless the
pump field itself has transverse wavevector components. Since the
$|{\mathbf q}|^{2}$ term in the expansion of $\Delta$, as given in
Eq.~(\ref{Delta-appr}), is smaller than $M{\bf e}_{2} \cdot {\bf
q}$ within the paraxial approximation, similar effects in type-I
quantum-interferometric measurements are expected to be smaller
given the same aperture size.

If the pump field is not a monochromatic planewave, but rather has
finite spectral bandwidth and transverse extent, the function
$\Delta$ defined after Eq.~(\ref{Phi-General}) can be approximated
by

\begin{equation}
\label{Delta-appr-spatial} \Delta \approx - D\nu + D_{p}\nu_{p} +
\frac{c}{\omega_{p}^{0}}\left[2|{\mathbf q}|^{2}+|\mathbf
q_{p}|^{2}\right] + M{\mathbf e}_{2}\cdot{\mathbf q} +
\left(M_{p}-{M\over 2}\right){\mathbf e}_{2}\cdot{\mathbf q}_{p}
\,,
\end{equation}

\noindent where
$D_{p}=\frac{1}{u_{p}}-\frac{1}{2}\left[\frac{1}{u_{o}}+\frac{1}{u_{e}}\right]$,
$u_{p}$ is the pump group velocity in the nonlinear medium,
$\nu_{p}=\omega_{p}-\omega_{p}^{0}$ is the deviation of the
angular pump frequency, ${\bf q}_{p}$ is the pump transverse
wavevector in the crystal, and $M_{p}$ is the spatial walk-off for
the pump beam. The second term in Eq.~(\ref{Delta-appr-spatial}),
which depends on the bandwidth of the pump field, is negligible if
the pump field is monochromatic or if $D_{p} \ll D$ for the
crystal under consideration. The last term in
Eq.~(\ref{Delta-appr-spatial}), which depends on the transverse
wavevector of the pump field, is negligible if the condition

\begin{equation}
\label{spatial-limit} {|(M_{p}-{M\over 2})|L \over a} \ll 1
\end{equation}

\noindent is satisfied, where $a$ is a characteristic width of the
pump beam at the crystal. For a 1.5-mm BBO crystal pumped by a
351-nm laser, $M=0.0711$ and $M_{p}=0.0770$, so that the pump
diameter would have to be less than or equal to 70 $\mu$m to
invalidate the planewave approximation. If such spatial walk-off
is neglected, the governing limit, arising from the remaining
contribution from diffraction [fourth term in
Eq.~(\ref{Delta-appr-spatial})], is found in
Ref.~\cite{Klyshko-Pump} to be about 10 $\mu$m for the
experimental arrangement presented in this work. Therefore, a
planewave approximation for the pump beam, whether continuous-wave
or pulsed, is valid for the experimental results presented in this
work, and indeed for most quantum-interference experiments.

 {\em Acknowledgments.---} This work was supported by the
National Science Foundation. The authors thank A.~F.~Abouraddy and
M.~C.~Booth for valuable suggestions.

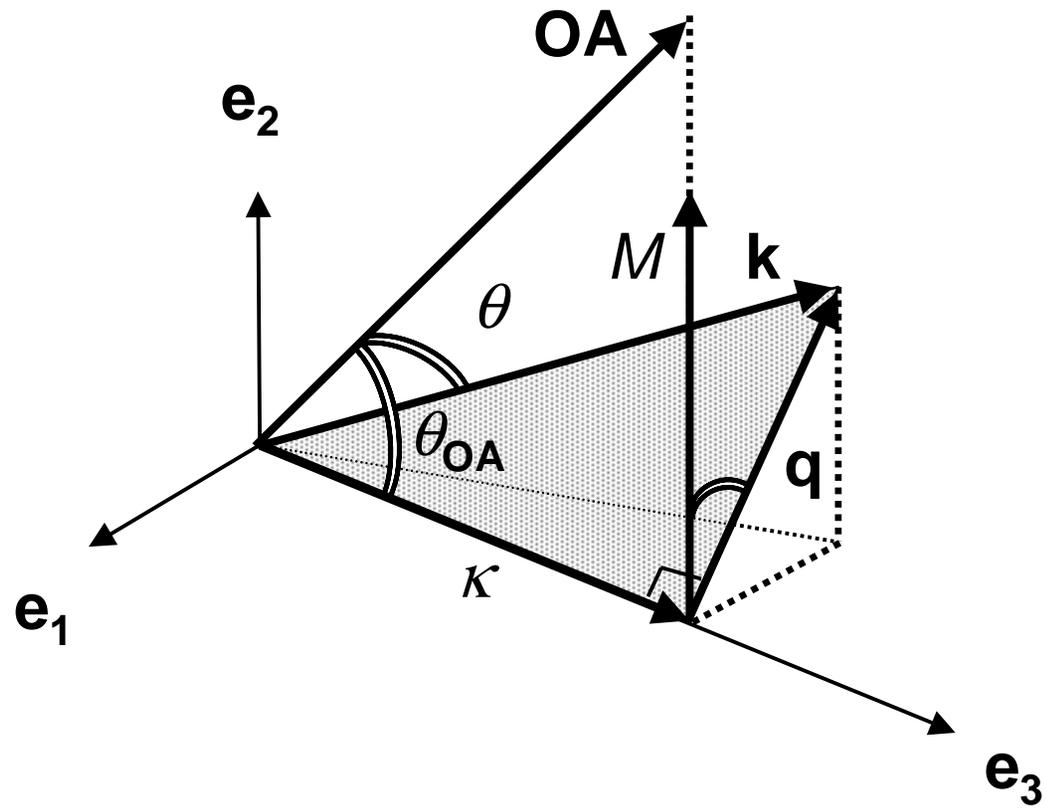
\begin{figure}
\label{Figure-1}
 \caption{Decomposition of a three-dimensional wavevector (${\bf k}$) into longitudinal
 (${\bf \kappa}$) and transverse (${\bf q}$) components. The angle between the
 optical axis of the nonlinear crystal (OA) and the wavevector ${\bf k}$ is $\theta$. The angle between
 the optical axis and the longitudinal axis (${\bf e}_{3}$) is denoted  ${\theta_{\rm OA}}$.
 The spatial walk-off of the extraordinary polarization component of a
 field travelling through the nonlinear crystal is characterized by the quantity $M$. } \label{autonum}
\end{figure}

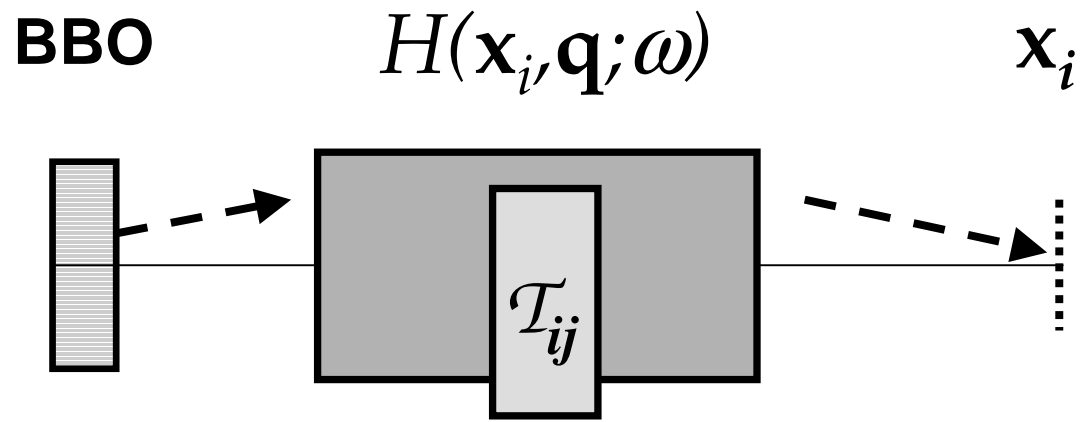
\begin{figure}
\label{Figure-2}
 \caption{(a) Illustration of the idealized setup
for observation of quantum interference using SPDC. BBO represents
a beta-barium borate nonlinear optical crystal, ${\cal
H}_{ij}({\bf x}_i, {\bf q}; \omega)$ is the transfer function of
the system, and the detection plane is represented by ${\bf x}_i$.
(b) For most experimental configurations the transfer function can
be factorized into diffraction-dependent $\left[H({\bf x}_i, {\bf
q}; \omega)\right]$ and diffraction-independent (${\cal T}_{ij}$)
components.} \label{autonum}
\end{figure}

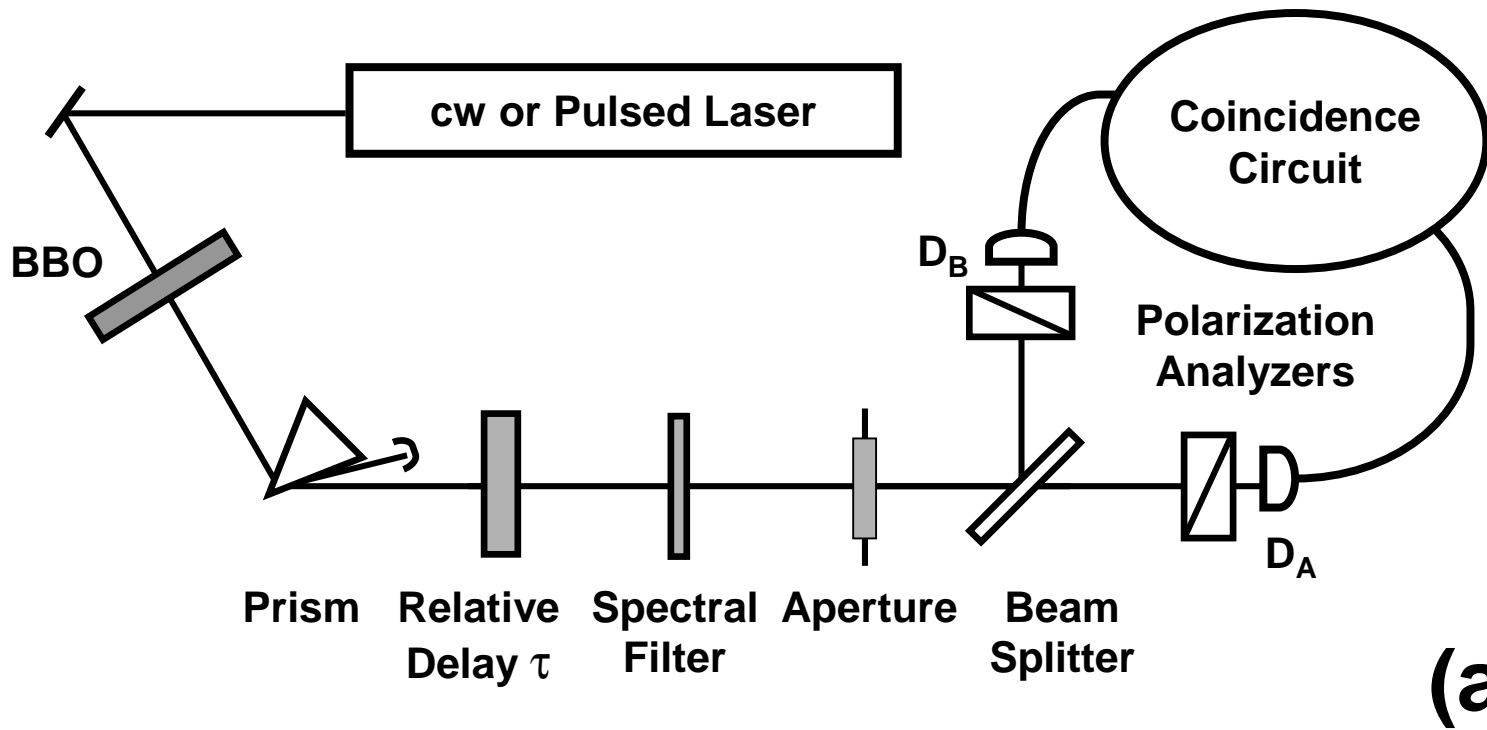
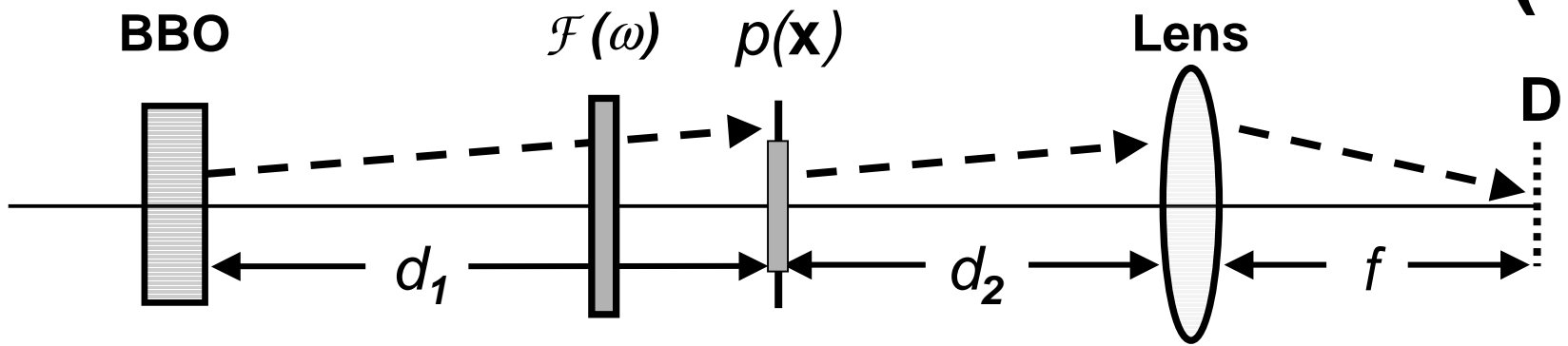
\begin{figure}
\label{Figure-3} \caption{(a) Schematic of the experimental setup
for observation of quantum interference using type-II collinear
SPDC (see text for details). (b) Detail of the path from the
crystal output plane to the detector input plane. ${\cal
F}(\omega)$ represents an (optional) filter transmission function,
$p({\bf x})$ represents an aperture function, and $f$ is the focal
length of the lens.} \label{autonum}
\end{figure}

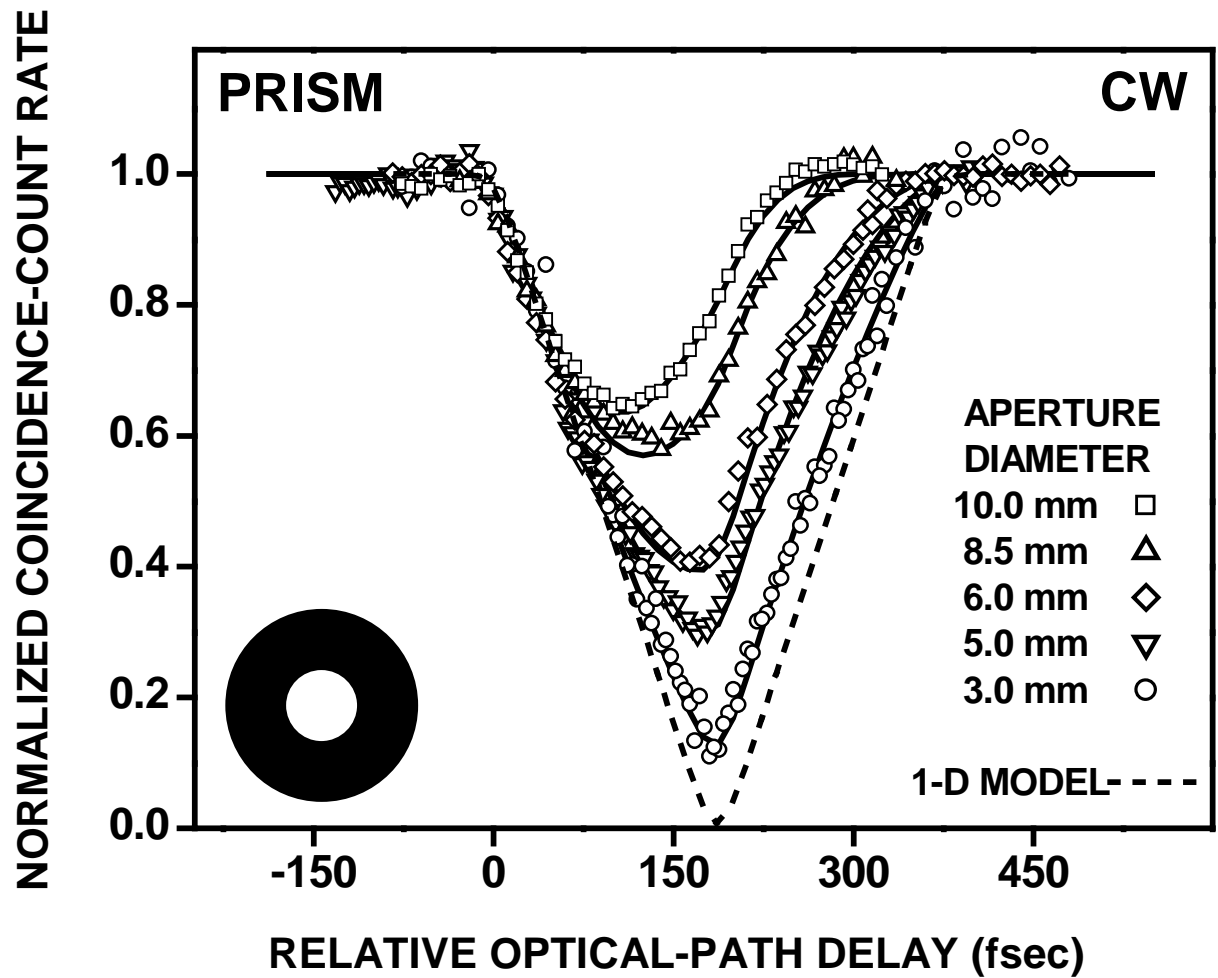
\begin{figure}
\label{Figure-4} \caption{Normalized coincidence-count rate
$R(\tau)/R_{0}$, as a function of the relative optical-path delay
$\tau$, for different diameters of a circular aperture placed 1~m
from the crystal. The symbols are the experimental results and the
solid curves are the theoretical plots for each aperture diameter.
The data was obtained using a 351-nm cw pump and no spectral
filters. No fitting parameters are used. The dashed curve
represents the one-dimensional (1-D) planewave theory.}
\label{autonum}
\end{figure}

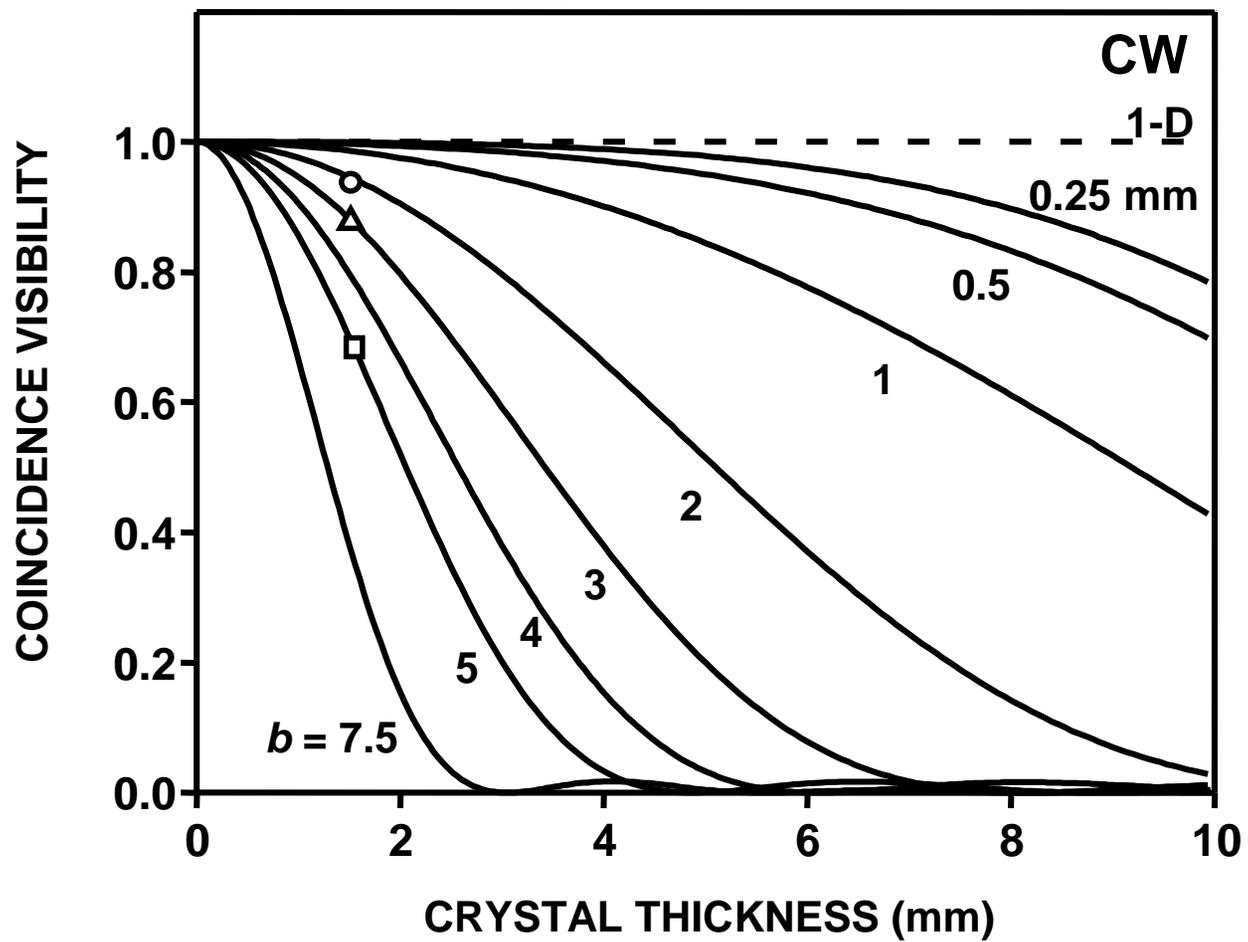
\begin{figure}
\label{Figure-5} \caption{Solid curves represent theoretical
coincidence visibility of the quantum-interference pattern for cw
SPDC as a function of crystal thickness, for various
circular-aperture diameters $b$ placed 1~m from the crystal.  The
dashed line represents the one-dimensional (1-D) planewave limit
of the multi-parameter formalism. Visibility is calculated for a
relative optical-path delay $\tau=LD/2$. The symbols represent
experimental data collected using a 1.5-mm-thick BBO crystal, and
aperture diameters $b$ of 2~mm (circle), 3~mm (triangle), and 5~mm
(square), as indicated on the plot.} \label{autonum}
\end{figure}

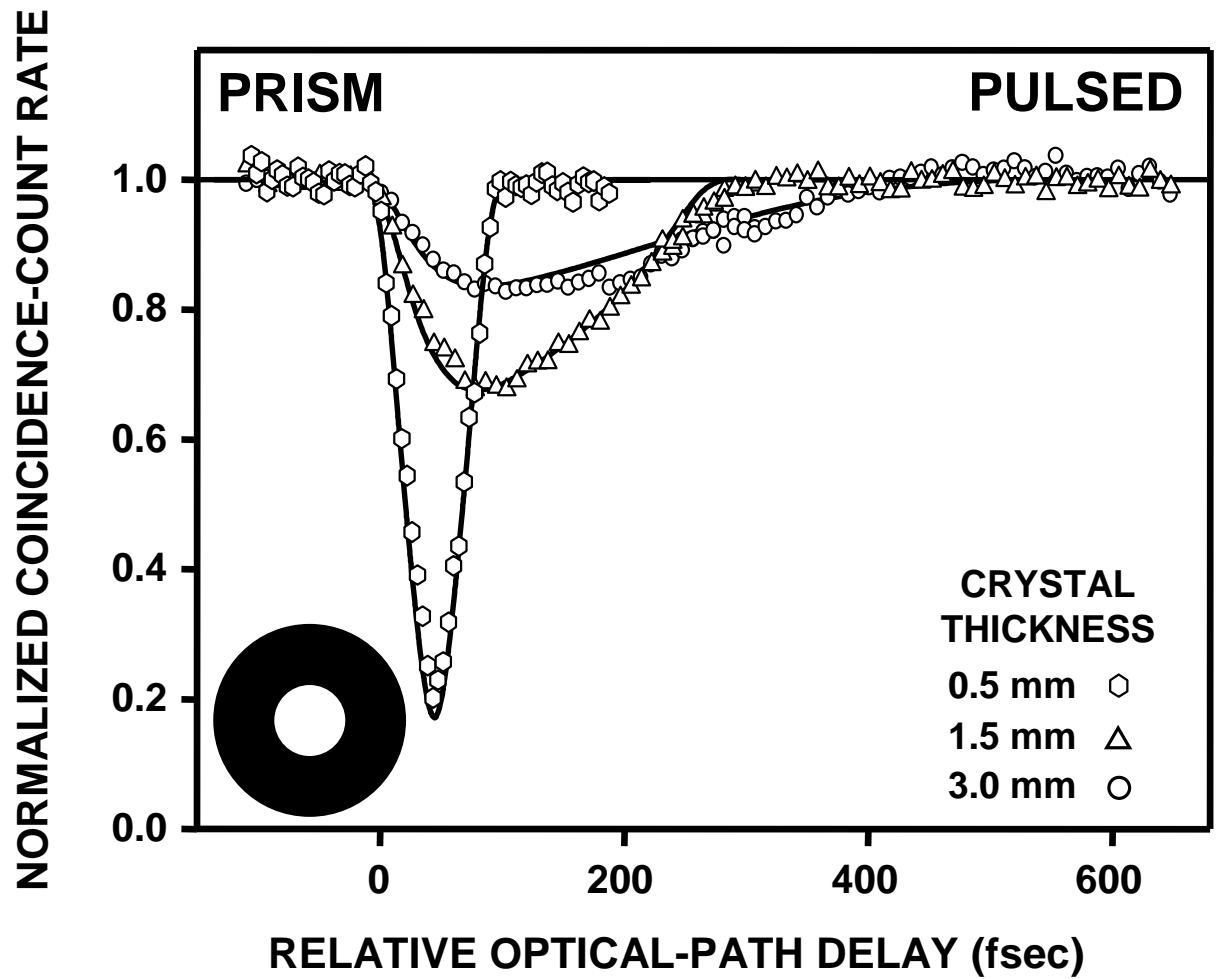
\begin{figure}
\label{Figure-6} \caption{Normalized coincidence-count rate, as a
function of the relative optical-path delay $\tau$, for 0.5-mm,
1.5-mm, and 3-mm thick BBO crystals as indicated in the plot. A
5-mm circular aperture was placed 1 m from the crystal. The
symbols are the experimental results and the solid curves are the
corresponding theoretical plots. The data were obtained using an
80-fsec pulsed pump centered at 415-nm and no spectral filters. No
fitting parameters are used.} \label{autonum}
\end{figure}

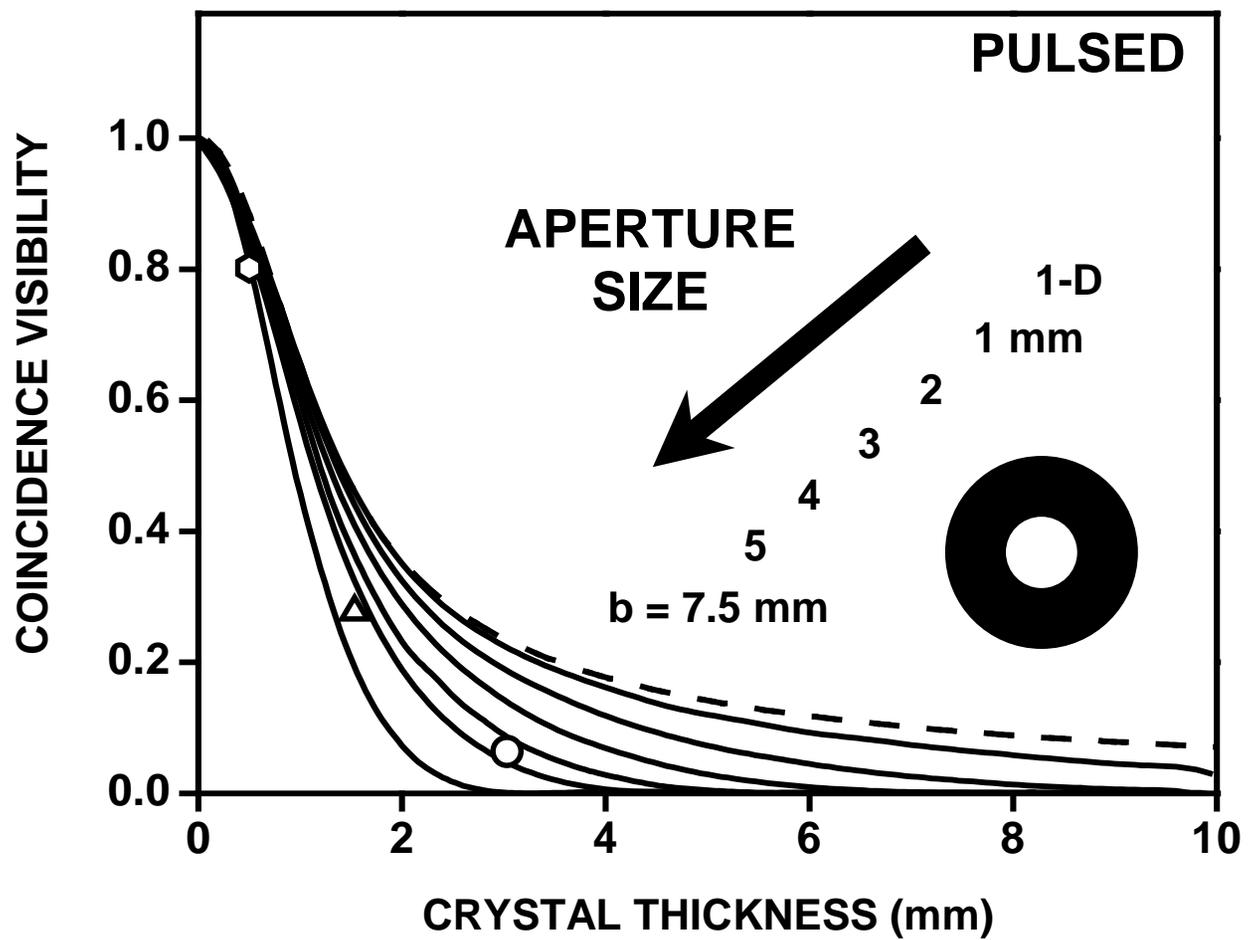
\begin{figure}
\label{Figure-7} \caption{Solid curves represent theoretical
coincidence visibility of the quantum-interference pattern for
pulsed SPDC as a function of crystal thickness, for various
circular-aperture diameters $b$ placed 1~m from the crystal.
Visibility is calculated for a relative optical-path delay
$\tau=LD/2$. The dashed curve represents the one-dimensional (1-D)
planewave limit of the multi-parameter formalism. The symbols
represent experimental data collected using a 3.0-mm aperture and
BBO crystals of thickness 0.5-mm (hexagon), 1.5-mm (triangle), and
3.0-mm (circle). Comparison should be made with Fig.~5 for the cw
SPDC. } \label{autonum}
\end{figure}

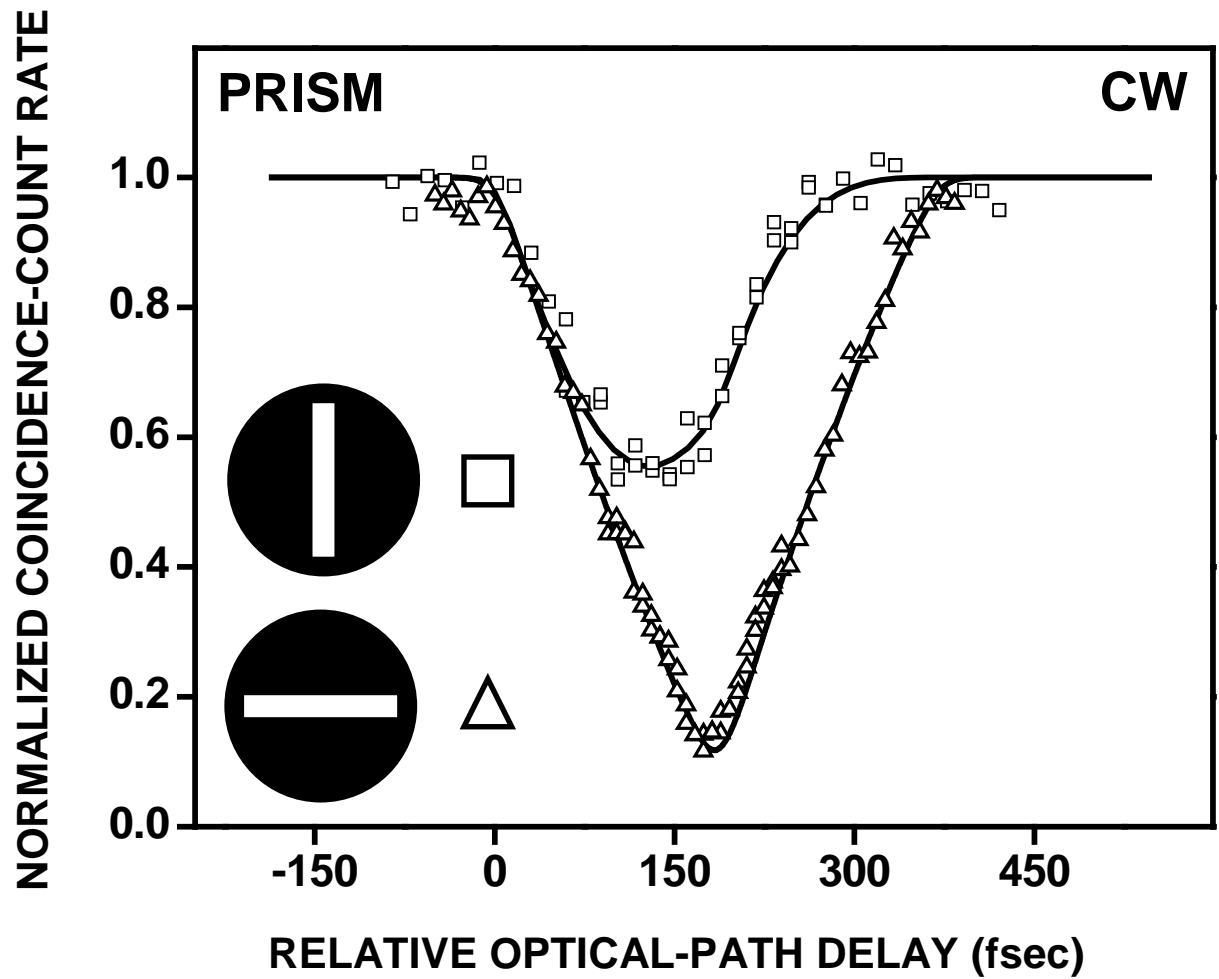
\begin{figure}
\label{Figure-8} \caption{Normalized coincidence-count rate as a
function of the relative optical-path delay for a $1\times 7$-mm
horizontal slit (triangles). The data was obtained using a 351-nm
cw pump and no spectral filters. Experimental results for a
vertical slit are indicated by squares. Solid curves are the
theoretical plots for the two orientations.} \label{autonum}
\end{figure}

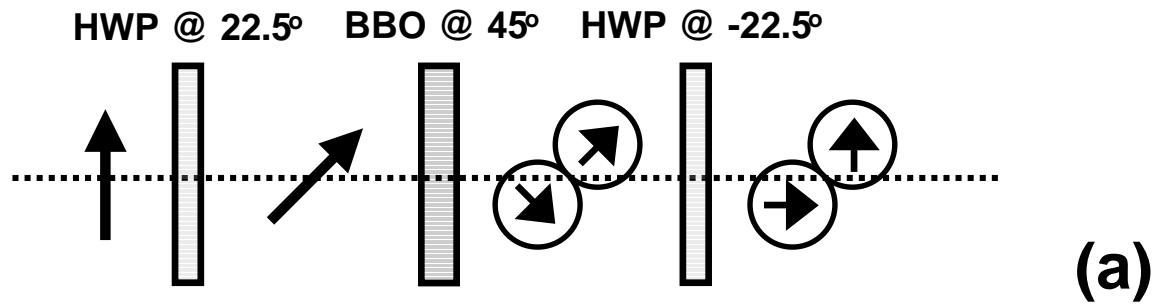
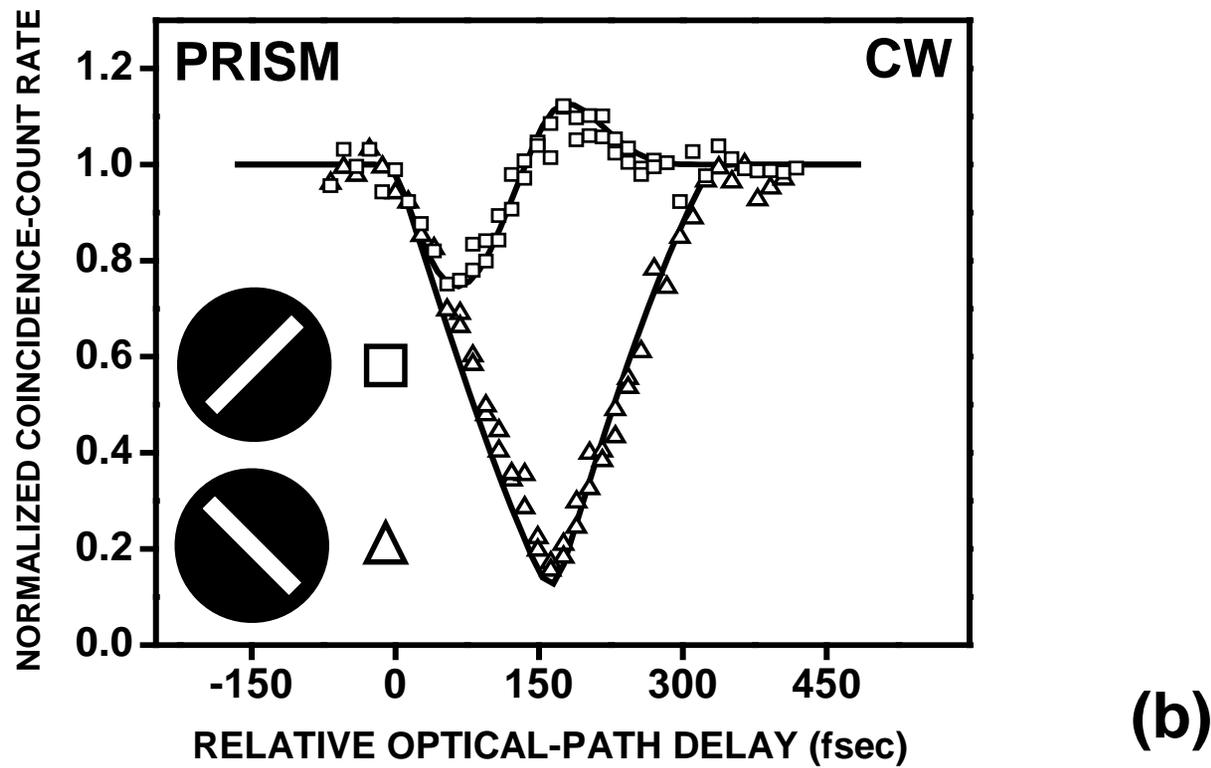
\begin{figure}
\label{Figure-9} \caption{(a) A half-wave plate at $22.5^\circ$
rotation, and another half-wave plate at $-22.5^\circ$ rotation
are placed before and after the BBO crystal in Fig.~3(a),
respectively. This arrangement results in SPDC polarized along the
horizontal-vertical axes and the axis of symmetry rotated
$45^\circ$ with respect to the vertical axis. (b) Normalized
coincidence-count rate from cw-pumped SPDC as a function of the
relative optical-path delay for a $1\times 7$-mm vertical slit
rotated $-45^\circ$ (triangles). Experimental results for a
vertical slit rotated $+45^\circ$ are indicated by squares. Solid
curves are the theoretical plots for the two orientations.}
\label{autonum}
\end{figure}

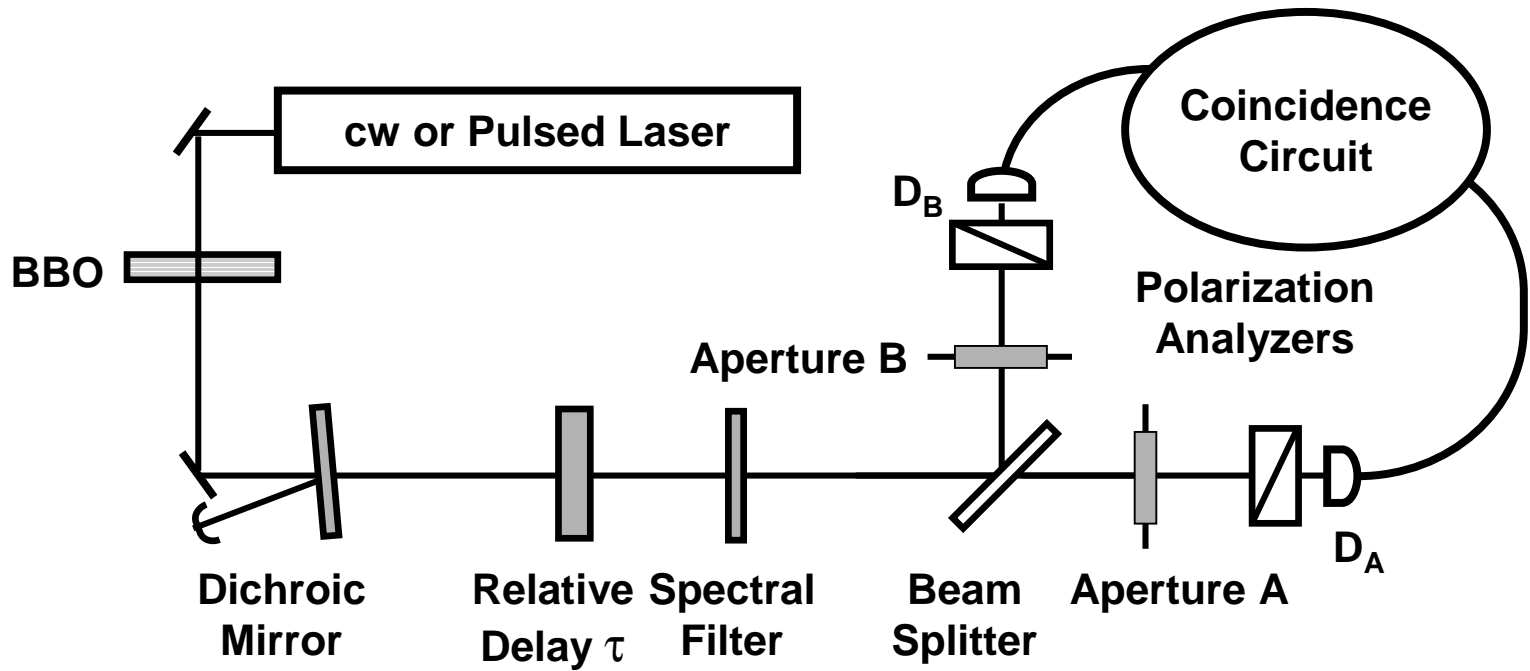
\begin{figure}
\label{Figure-10} \caption{Schematic of alternate experimental
setup for observation of quantum interference using cw-pumped
type-II collinear SPDC. The configuration illustrated here makes
use of a dichroic mirror in place of the prism used in Fig.~3(a),
thereby admitting greater acceptance of the transverse-wave
components. The dichroic mirror reflects the pump wavelength while
transmitting a broad wavelength range that includes the bandwidth
of the SPDC. The single aperture shown in Fig. 3(a) is replaced by
seperate apertures placed equal distances from the beamsplitter in
each arm of the interferometer.} \label{autonum}
\end{figure}

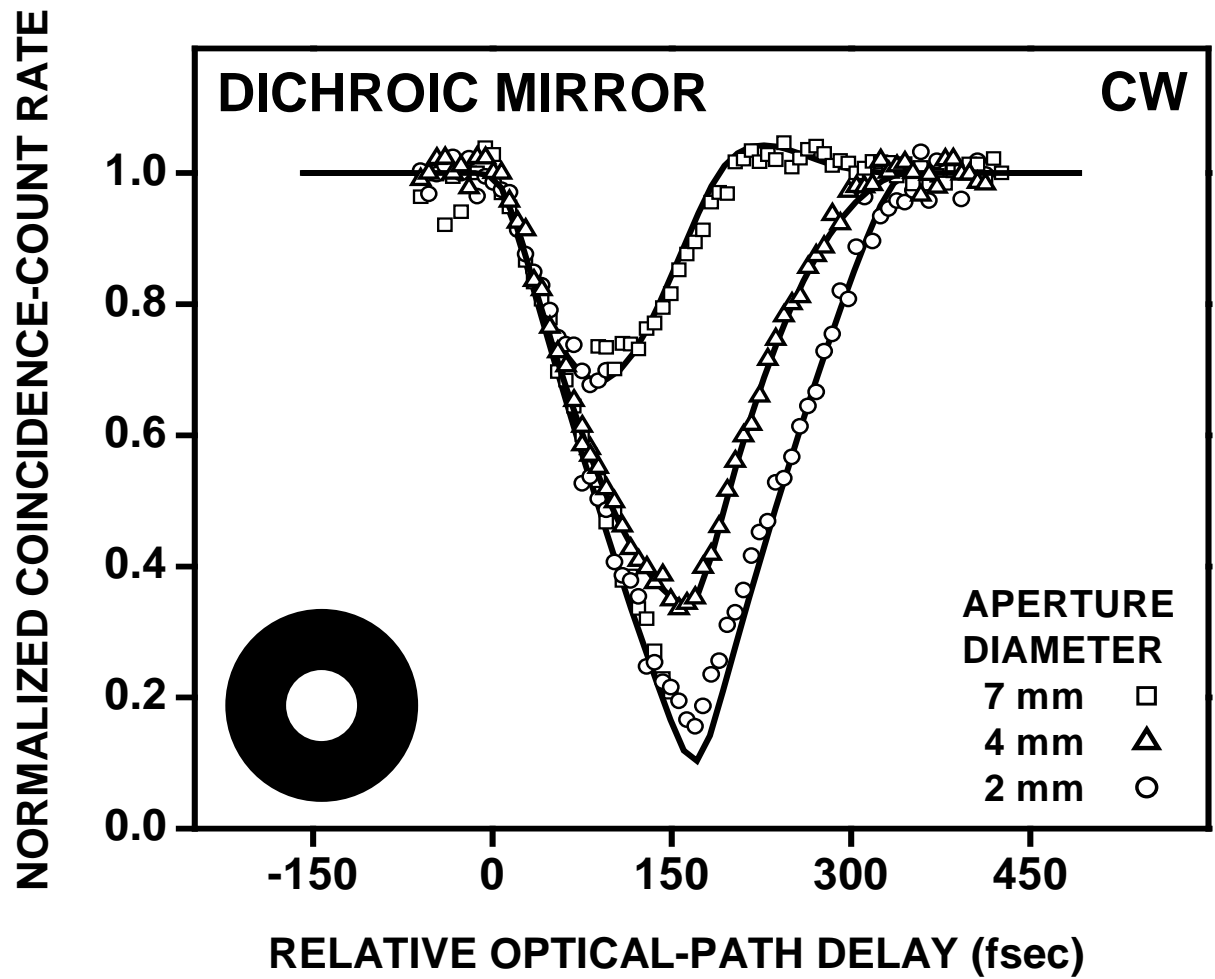
\begin{figure}
\label{Figure-11} \caption{Normalized coincidence-count rate as a
function of the relative optical-path delay $\tau$, for different
diameters of an aperture that is circular in the configuration of
Fig.~10. The symbols are the experimental results and the solid
curves are the theoretical plots for each aperture diameter. The
data were obtained using a 351-nm cw pump and no spectral filters.
No fitting parameters are used. The behavior of the interference
pattern is similar to that observed in Fig.~4; the dependence on
the diameter of the aperture is slightly stronger in this case.}
\label{autonum}
\end{figure}

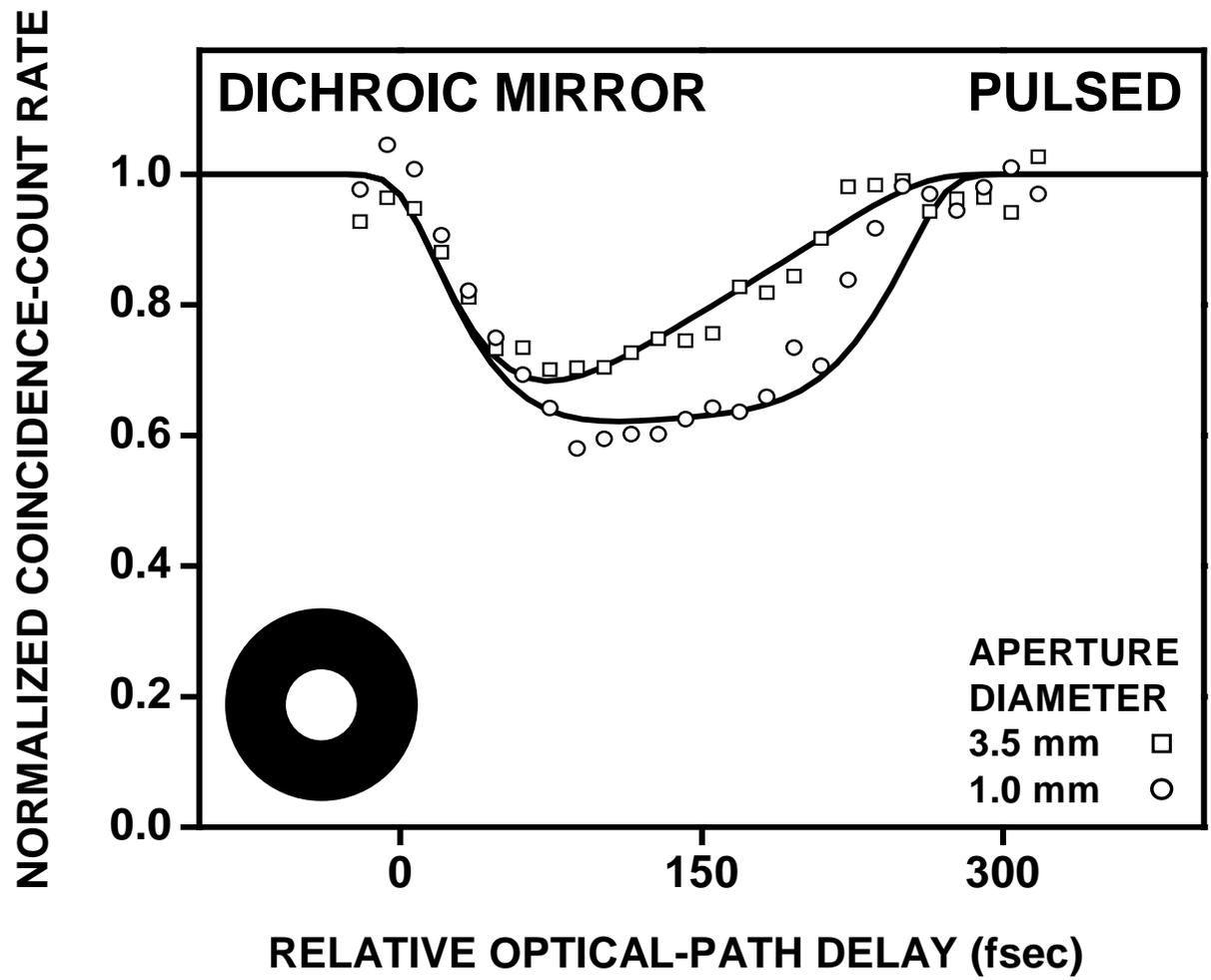
\begin{figure}
\label{Figure-12} \caption{Normalized coincidence-count rate as a
function of the relative optical-path delay, for different
diameters of an aperture that is circular in the configuration of
Fig.~10. The symbols are the experimental results and the solid
curves are the theoretical plots for each aperture diameter. The
data were obtained using an 80-fsec pulsed pump centered at 415-nm
and no spectral filters. No fitting parameters are used.
Comparison should be made with Fig.~11 for the cw SPDC case.}
\label{autonum}
\end{figure}

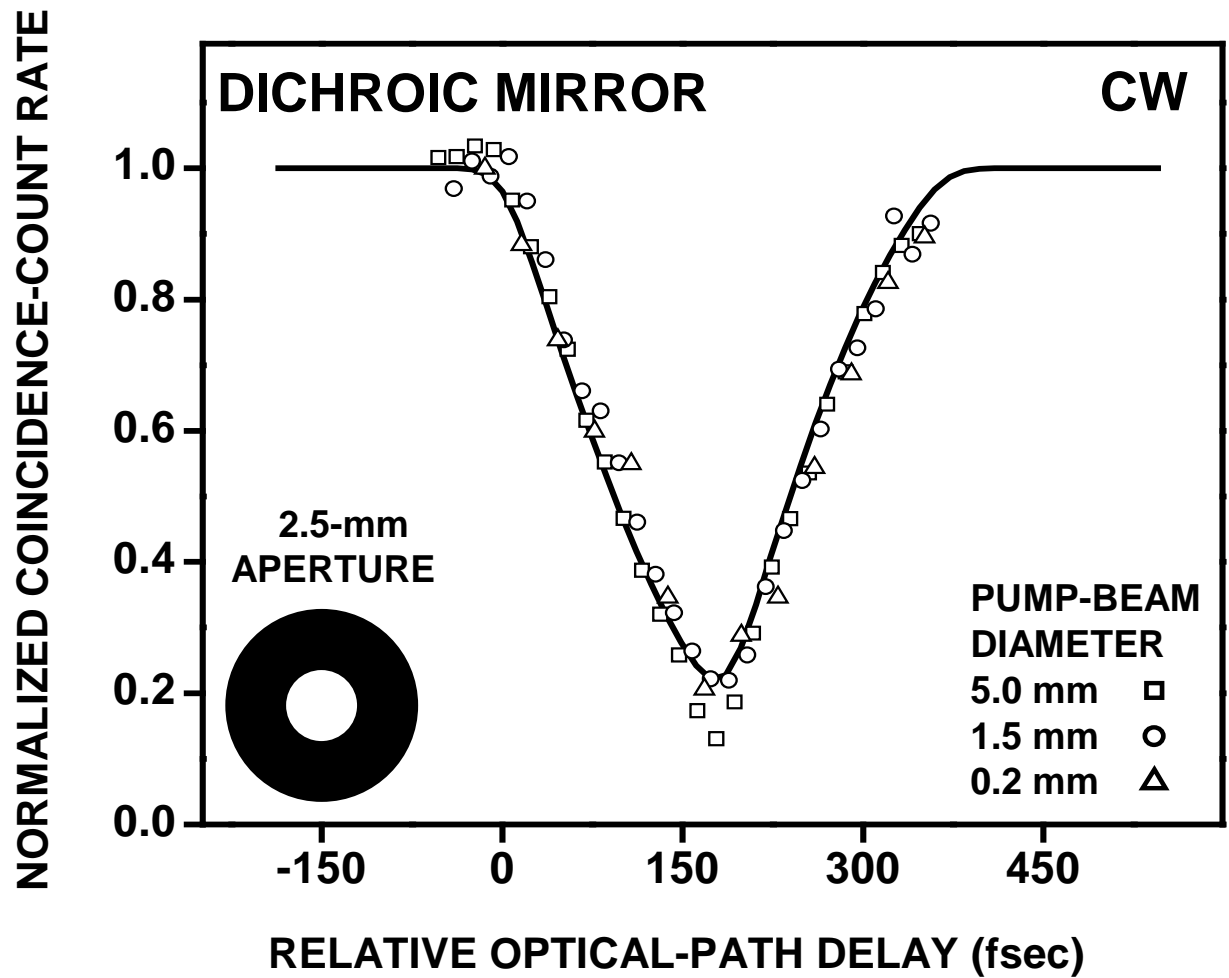
\begin{figure}
\label{Figure-13} \caption{Normalized coincidence-count rate as a
function of the relative optical-path delay, for different
diameters of the pump beam in the configuration of Fig.~10. The
symbols are the experimental results and the solid curve is the
theoretical plot of the quantum-interference pattern for an
infinite planewave pump. The data were obtained using a 351-nm cw
pump and no spectral filters. The circular aperture in the optical
system for the down-converted light was 2.5~mm at a distance of
750~mm. No fitting parameters are used.} \label{autonum}
\end{figure}

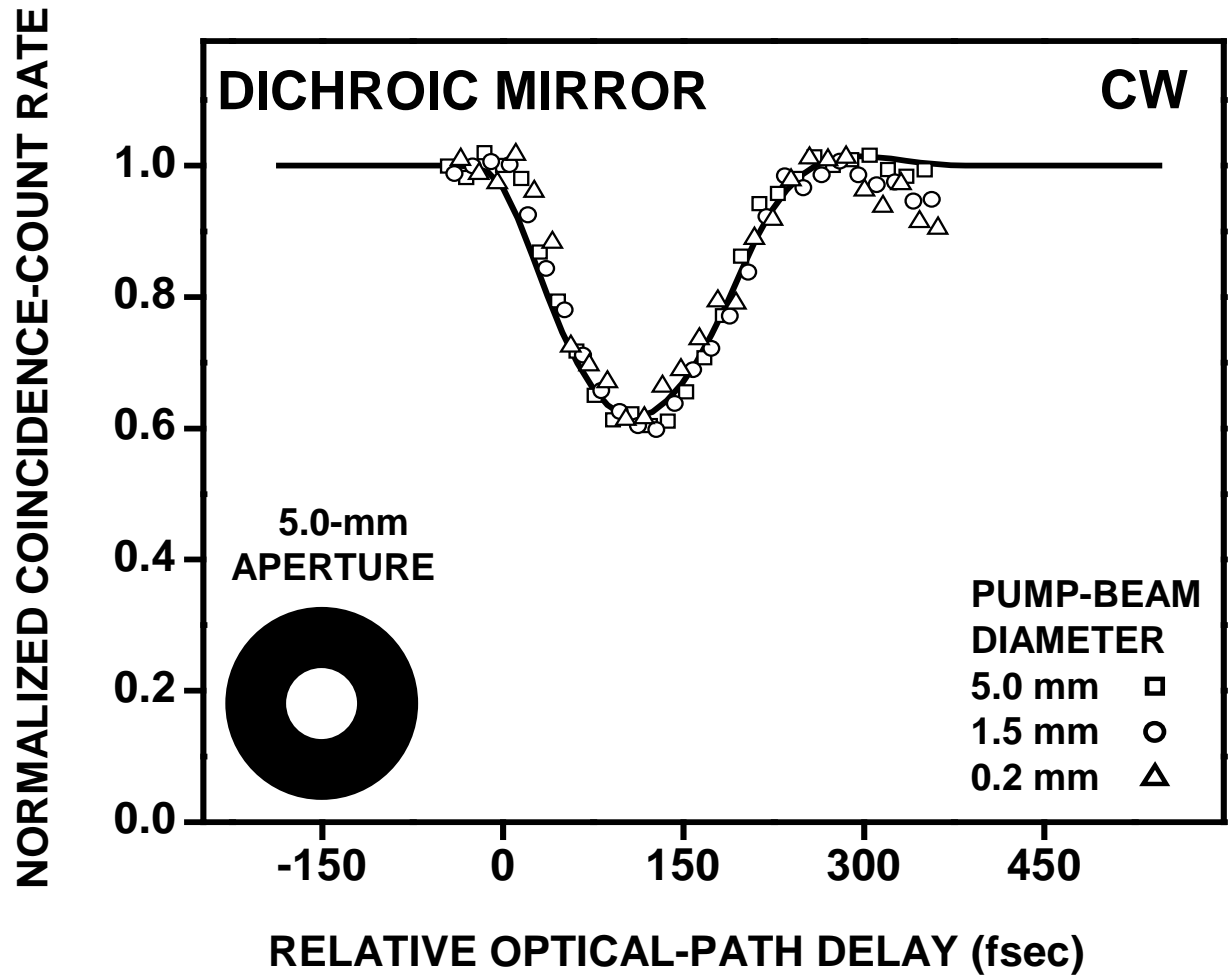
\begin{figure}
\label{Figure-14} \caption{Plots similar to those in Fig.~13 in
the presence of a 5.0-mm circular aperture in the optical system
for the down-converted light. } \label{autonum}
\end{figure}

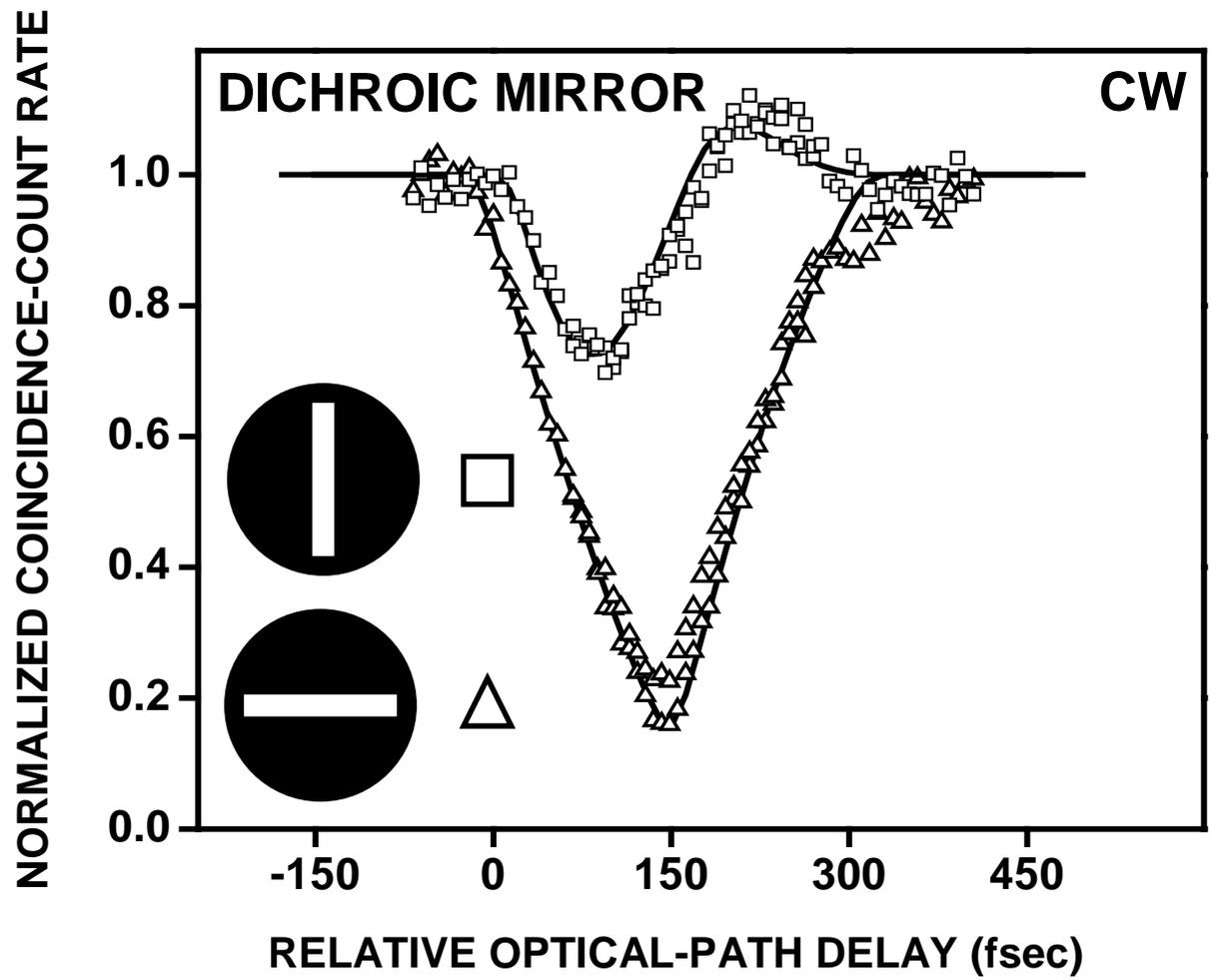
\begin{figure}
\label{Figure-15} \caption{Normalized coincidence-count rate as a
function of the relative optical-path delay for identical $1\times
7$-mm horizontal slits placed in each arm in the configuration of
Fig.~10 (triangles). The data were obtained using a 351-nm cw pump
and no spectral filters. Experimental results are also shown for
two identical $1\times 7$-mm vertical slits, but shifted with
respect to each other by 1.6~mm along the long axis of the slit
(squares). Solid curves are the theoretical plots for the two
orientations.} \label{autonum}
\end{figure}

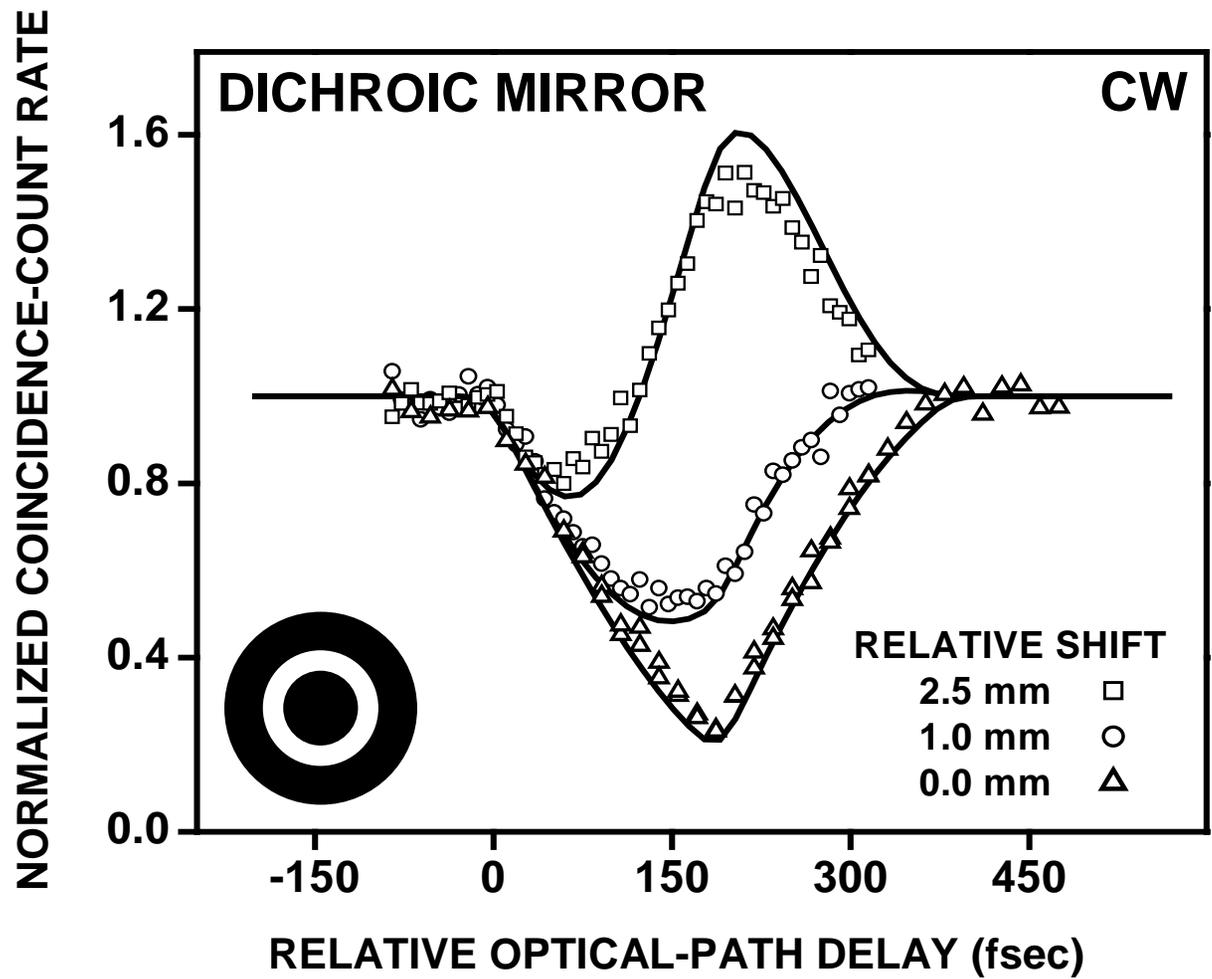
\begin{figure}
\label{Figure-16} \caption{Normalized coincidence-count rate as a
function of the relative optical-path delay, for an annular
aperture (internal and external diameters of 2 and 4~mm,
respectively) in one of the arms of the interferometer in the
configuration of Fig.~10. A 7-mm circular aperture is placed in
the other arm. The data were obtained using a 351-nm cw pump and
no spectral filters. The symbols are experimental results for
different relative shifts of the annulus along the direction of
the optical axis of the crystal (vertical). The solid curves are
the theoretical plots without any fitting parameters. }
\label{autonum}
\end{figure}

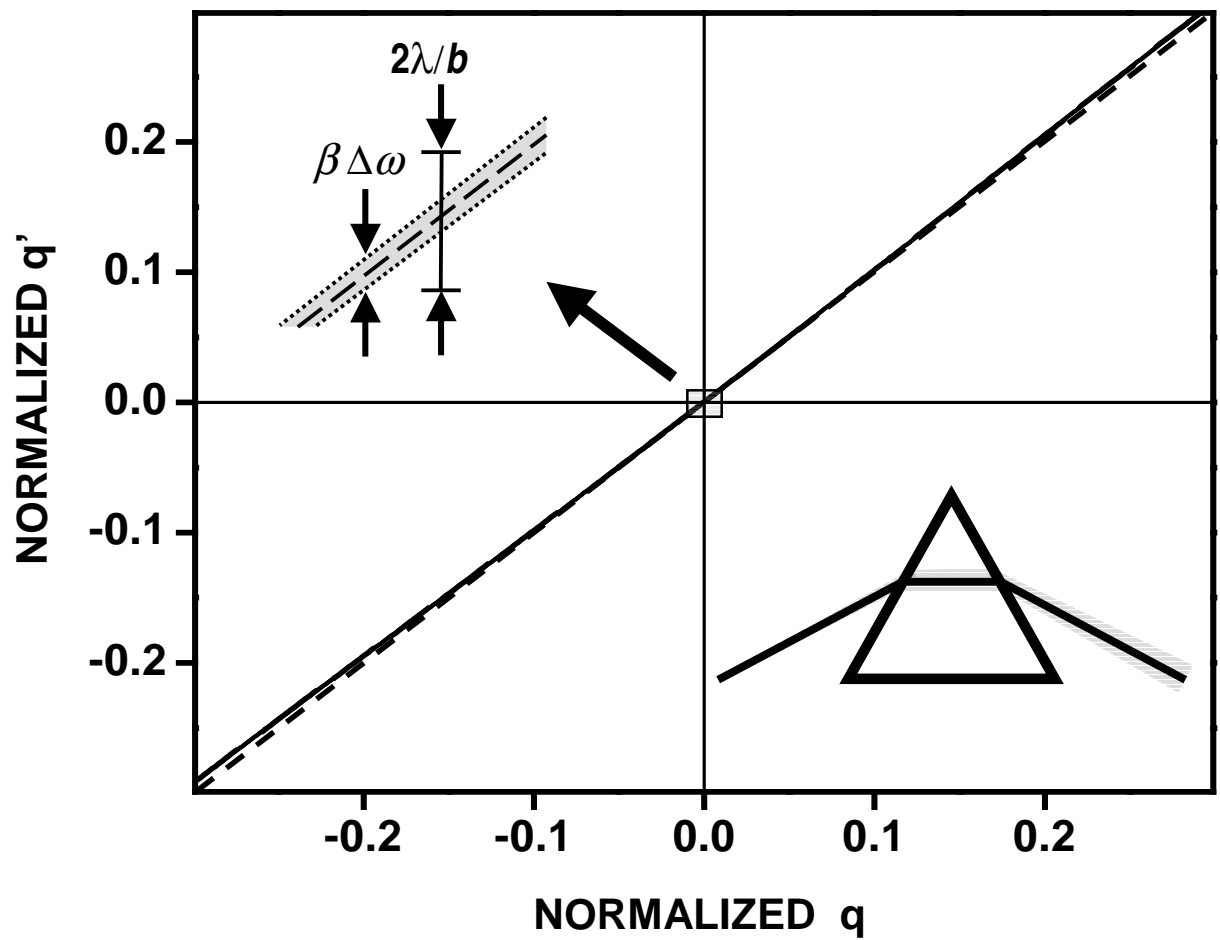
\begin{figure}
\label{Figure-17} \caption{A plot of Eq.~(A2), the relation
between the magnitudes of the transverse wavevectors entering
(${\bf q}$) and exiting (${\bf q'}$) a f used-silica prism with an
apex angle of $60^\circ$ (see inset at lower right). The range of
transverse wavevectors allowed by the optical system in our
experiments lies completely within the area included in the tiny
box at the center of the plot. The upper left inset shows details
of this central region. The dashed curve is a line with unity
slope representing the $|{\bf q'}|=|{\bf q}|$ map. The dotted
lines denote the deviation from this mapping arising from the
spectral bandwidth ($\Delta\omega$) of the incident light beam. As
long as the width of the shaded area, which denotes the angular
dispersion of the prism, is smaller than the angular resolution
($2\lambda/b$) of the aperture-lens combination, the effect of the
prism is negligible.}
\end{figure}

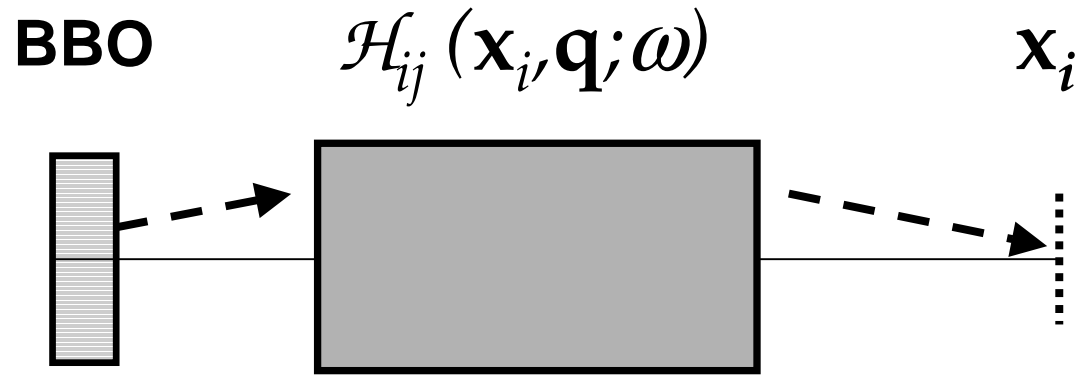
\end{document}